\documentclass[10pt]{article}

\usepackage{latexsym}
\usepackage[dvips]{epsfig} 
\usepackage{amssymb,amsmath,amsthm,mathrsfs} 
\usepackage{graphicx}
\usepackage{lineno}

\usepackage[authoryear,sort &compress]{natbib}
\usepackage{url}
\usepackage{enumerate}
\usepackage{colordvi,color,pspicture}
\usepackage{psfrag}
\usepackage{rotating}
\usepackage{algorithmic}

\newcommand{\al}{\alpha}
\newcommand{\be}{\beta}
\newcommand{\ah}{\widehat{\alpha}}
\newcommand{\bh}{\widehat{\beta}}
\newcommand{\ve}{\varepsilon}
\newcommand{\E}{\mathbf{E}}
\newcommand{\U}{\mathcal{U}}
\newcommand{\I}{\mathbf{I}}

\theoremstyle{plain}

\theoremstyle{definition}

\theoremstyle{remark}

\begin{document}

\title{Technical Report \# KU-EC-09-4:\\
A Comparison of Analysis of Covariate-Adjusted Residuals and Analysis of Covariance}
\author{
E. Ceyhan$^{1^\ast}$, Carla L. Goad$^{2}$
}

\date{\today}

\maketitle

\begin{center}
$^{1}$\textit{Department of Mathematics, Ko\c{c} University, 34450, Sar{\i}yer, Istanbul, Turkey.}\\
$^{2}$\textit{Department of Statistics, Oklahoma State University, Stillwater, OK 74078-1056, USA.}
\end{center}

\noindent
*\textbf{corresponding author:}\\
Elvan Ceyhan,\\
Dept. of Mathematics, Ko\c{c} University,\\
Rumelifeneri Yolu, 34450 Sar{\i}yer,\\
Istanbul, Turkey\\
{\bf e-mail:} elceyhan@ku.edu.tr\\
{\bf phone:} +90 (212) 338-1845\\
{\bf fax:} +90 (212) 338-1559\\

\noindent
\textbf{short title:} ANOVA on Covariate-Adjusted Residuals and ANCOVA

\begin{abstract}
\noindent
Various methods to control the influence of a covariate on
a response variable are compared.
In particular,
ANOVA with or without homogeneity of variances (HOV) of errors
and Kruskal-Wallis (K-W) tests on covariate-adjusted residuals and
analysis of covariance (ANCOVA) are compared.
Covariate-adjusted residuals are obtained from the overall
regression line fit to the entire data set ignoring the treatment levels or factors.
The underlying assumptions for ANCOVA and methods on covariate-adjusted residuals
are determined and the methods are compared only when both methods
are appropriate.
It is demonstrated that the methods on covariate-adjusted residuals
are only appropriate in removing the covariate influence
when the treatment-specific lines are parallel and
treatment-specific covariate means are equal.
Empirical size and power performance of the methods are compared
by extensive Monte Carlo simulations.
We manipulated the conditions such as assumptions of
normality and HOV, sample size, and clustering of the covariates.
The parametric methods (i.e., ANOVA with or without HOV on
covariate-adjusted residuals and ANCOVA) exhibited similar size and power
when error terms have symmetric distributions with variances having
the same functional form for each treatment,
and covariates have uniform distributions within the same interval for each treatment.
For large samples, it is shown that the
parametric methods will give similar results if sample covariate means for
all treatments are similar.
In such cases, parametric tests have higher
power compared to the nonparametric K-W test on covariate-adjusted residuals.
When error terms have asymmetric distributions or have variances that are heterogeneous with different
functional forms for each treatment,
ANCOVA and analysis of covariate-adjusted residuals are liberal
with K-W test having higher power than the parametric tests.
The methods on covariate-adjusted residuals
are severely affected by the clustering of the covariates relative to the treatment factors,
when covariate means are very different for treatments.
For data clusters, ANCOVA method exhibits the appropriate level.
However such a clustering might suggest dependence between the covariates
and the treatment factors, so makes ANCOVA less reliable as well.
Guidelines on which method to use for various cases are also provided.
\end{abstract}



\newpage


\section{Introduction}
\label{sec:intro}
In an experiment, the response variable may depend
on the treatment factors and quite often on some external factor
that has a strong influence on the response variable.
If such external factors are qualitative or discrete,
then {\it blocking} can be performed to remove their influence.
However, if the external factors are quantitative and continuous,
the effect of the external factor can be
accounted for by adopting it as a {\it covariate} (\cite{kuehl:2000}),
which is also called a {\it concomitant variable} (\cite{ott:1993}, \cite{milliken:2002}).
Throughout this article, a covariate is defined
to be a variable that may affect the relationship between the
response variable and factors (or treatments) of interest,
but is not of primary interest itself.
\cite{maxwell:1984} compared methods of incorporating a covariate
into an experimental design
and showed that it is not correct to consider the correlation
between the dependent variable and covariate in choosing the best technique.
Instead, they recommend considering whether scores
on the covariate are available for all subjects prior to
assigning any subject to treatment conditions and
whether the relationship of the dependent variable and covariate is linear.

In various disciplines such as ecology, biology, medicine, etc.
the goal is comparison of a response variable among several treatments
after the influence of the covariate is removed.
Different techniques are used or suggested in statistical and biological literature
to remove the influence of the covariate(s) on the response variable (\cite{huitema:1980}).
For example in ecology,
one might want to compare richness-area relationships among regions,
shoot ratios of plants among several treatments,
and of C:N ratios among sites (\cite{garcia-berthou:2001}).\
There are three main statistical techniques for attaining that goal:
(i) analysis of the ratio of response to the covariate;
(ii) analysis of the residuals from the regression of the response with the covariate;
and
(iii) analysis of covariance (ANCOVA).

Analysis of the ratios is perhaps the oldest method used to remove the covariate effect
(e.g., size effect in biology) (see \cite{albrecht:1993} for a comprehensive review).
Although many authors recommend its disuse (\cite{packard:1988}, \cite{atchley:1976}),
it might still appear in literature on occasion (\cite{albrecht:1993}).
For instance, in physiological and nutrition research,
the data are scaled by taking the ratio of the response variable to the covariate.
Using the ratios in removing the influence of the covariate on the response depends on
the relationship between the response and the covariate variables (\cite{raubenheimer:1992}).
Regression analysis of a response variable on the covariate(s) is
common to detect such relationships,
which are categorized as {\it isometric} or {\it allometric} relationships (\cite{small:1996}).
{\it Isometry} occurs when the relationship between a response variable and the
covariate is linear with a zero intercept.
If the relationship is nonlinear or if there is a non-zero intercept,
it is called {\it allometry}.
In allometry, the influence of the covariate cannot be removed
by taking the ratio of the response to the covariate.
In both of allometry and isometry cases,
ANOVA on ratios (i.e., {\it response}/{\it covariate} values)
introduces heterogeneity of variances in the error terms which violates
an assumption of ANOVA (with homogeneity of variances (HOV)).
Hence, ANOVA on ratios may give spurious and invalid results for treatment comparisons,
so ANCOVA is recommended over the use of ratios (\cite{raubenheimer:1992}).
See \cite{ceyhan:masters-thesis} for a detailed discussion on the
use of ratios to remove the covariate influence.

An alternative method to remove the effect of a covariate on the response variable
in biological and ecological research is the use of residuals (\cite{garcia-berthou:2001}).
In this method an overall regression line is fitted to the entire data set and
residuals are obtained from this line (\cite{beaupre:1998}).
These residuals will be referred to as {\it covariate-adjusted residuals}, henceforth.
This method was recommended in ecological literature by \cite{jakob:1996}
who called it ``residual index" method.
Then treatments are compared with ANOVA with HOV on these residuals.

Due to the problems associated with the use of ratios in removing the
influence of the covariate from the response,
ANOVA (with HOV) on covariate-adjusted residuals and ANCOVA were
recommended over the use of ratios (\cite{packard:1988} and \cite{atchley:1976}).
For example, \cite{beaupre:1998} used ANOVA on
covariate-adjusted residuals in a zoological study.
\cite{ceyhan:masters-thesis} compared the ANCOVA and
ANOVA (with HOV) on covariate-adjusted residuals.
ANCOVA has been widely applied in ecology
and it was shown to be a superior alternative to ratios by \cite{garcia-berthou:2001}
who also point out problems with the residual index and
recommends ANCOVA as the correct alternative.
They also discuss the differences between ANCOVA and ANOVA on the residual index.
They argue that the residual analysis is totally misleading as
(i) residuals are obtained from an overall regression on the pooled data,
(ii) the residual analysis uses the wrong degrees of freedom in inference,
and (iii) residuals fail to satisfy the ANOVA assumptions even if
the original data did satisfy them.
In fact,
\cite{maxwell:1985} also demonstrated the inappropriateness of ANOVA on residuals.

Although ANCOVA is a well-established and highly recommended tool,
it also has critics.
However, the main problem in literature is not the inappropriateness of ANCOVA,
rather its misuse and misinterpretation.
For example, \cite{rheinheimer:2001} investigated how
the empirical size and power performances of ANCOVA are affected
when the assumptions of normality and HOV, sample size, number of treatment
groups, and strength of the covariate-dependent variable
relationship are manipulated.
They demonstrated that for balanced designs,
the ANCOVA $F$ test was robust and was often the most
powerful test through all sample-size designs and distributional configurations.
Otherwise it was not the best performer.
In fact, the assumptions for ANCOVA are crucial for its use;
especially, the independence between the covariate and the treatment
factors is an often ignored assumption resulting incorrect inferences (\cite{miller-ANCOVA:2001}).
This violation is very common in fields such as psychology and psychiatry,
due to nonrandom group assignment in observational studies,
and \cite{miller-ANCOVA:2001} also suggest some alternatives for such cases.
Hence the recommendations in favor on ANCOVA (including ours)
are valid only when the underlying assumptions are met.

In this article,
we demonstrate that it is not always wrong to use the residuals.
We also discuss the differences between ANCOVA and analysis of residuals,
provide when and under what conditions the two procedures are
appropriate and comparable.
Then under such conditions, we not only consider ANOVA (with HOV),
but also ANOVA without HOV and Kruskal-Wallis (K-W) test on the covariate-adjusted residuals.
We provide the empirical size performance of each method under the null case
and the empirical power under various alternatives
using extensive Monte Carlo simulations.

The nonparametric analysis by K-W test on the covariate-adjusted
residuals is actually not entirely nonparametric,
in the sense that, the residuals are obtained from a fully parametric model.
However, when the covariate is not continuous but categorical
data with ordinal levels, then a nonparametric version of ANCOVA
can be performed (see, e.g., \cite{akritas:2000} and \cite{tsangari:2004JMVA}).
Further, the nonparametric ANCOVA model of \cite{akritas:2000} is extended to longitudinal
data for up to three covariates (\cite{tsangari:2004JNPS}).
Additionally, there are nonparametric methods such as Quade's procedure, Puri
and Sen's solution, Burnett and Barr's rank difference scores,
Conover and Iman's rank transformation test, Hettmansperger's
procedure, and the Puri-Sen-Harwell-Serlin test
which can be used as alternatives to ANCOVA (see \cite{rheinheimer:2001}
for the comparison of the these tests with ANCOVA and relevant references).
In fact, \cite{rheinheimer:2001} showed that with unbalanced designs, with variance
heterogeneity, and when the largest treatment-group variance was
matched with the largest group sample size,
these nonparametric alternatives generally outperformed the ANCOVA test.

The methods to remove covariate influence on the response are presented in
Section \ref{sec:ANCOVA-ANOVA-Cov-Adj-Res}, where the ANCOVA method,
ANOVA with HOV and without HOV on covariate adjusted residuals,
and K-W test on covariate-adjusted residuals are described.
A detailed comparison of the methods, in terms of the null hypotheses,
and conditions under which they are equivalent are provided
in Section \ref{sec:comparison}.
The Monte Carlo simulation analysis used for the comparison of the methods
in terms of empirical size and power is provided in Section \ref{sec:MonteCarlo}.
A discussion together with a detailed guideline on
the use of the discussed methods is provided in Section \ref{sec:disc-conc}.

\section{ANCOVA and Methods on Covariate-Adjusted Residuals}
\label{sec:ANCOVA-ANOVA-Cov-Adj-Res}
In this section, the models and the corresponding assumptions for ANCOVA and
the methods on covariate-adjusted residuals are provided.

\subsection{ANCOVA Method}
\label{sec:ANCOVA}
For convenience, only ANCOVA with a one-way treatment
structure in a completely randomized design and
a single covariate is investigated.
A simple linear relationship between the covariate and the
response for each treatment level is assumed.

Suppose there are $t$ levels of a treatment factor,
with each level having $s_i$ observations;
and there are $r_{ij}$ replicates for each covariate value for treatment
level $i$ for $i=1,2,\ldots,t$ and $j=1,2,\ldots,n_i$
where $n_i$ is the number of distinct covariate values at treatment level $i$.
Let $n$ be the total number of observations in the entire data set
then $s_i=\sum_{j=1}^{n_i}{r_{ij}}$ and $n=\sum_{i=1}^t{s_i}$.
ANCOVA fits a straight line to each treatment level.
These lines can be modeled as
\begin{equation}
\label{eqn:trt-spec-lines}
Y_{ijk} \,=\,\mu_i+\be_i \,X_{ij}+e_{ijk}
\end{equation}
where $X_{ij}$ is the $j^{th}$ value of the covariate for treatment level $i$,
$Y_{ijk}$ is the $k^{th}$ response at $X_{ij}$,
$\mu_i$ is the intercept and $\be_i$ is the slope for treatment level $i$,
and $e_{ijk}$ is the random error term for
$i=1,2,\ldots,t$, $j=1,2,\ldots,n_i$, and $k=1,2,\ldots,r_{ij}$.
The assumptions for the ANCOVA model in Equation \eqref{eqn:trt-spec-lines} are:
(a) The $X_{ij}$ (covariate) values are assumed to be fixed as in regression analysis
(i.e., $X_{ij}$ is not a random variable).
(b) $e_{ijk} \stackrel{iid}{\sim}N\left({0,\sigma_e^2} \right)$ for all treatments where
$\stackrel{iid}{\sim}$ stands for ``independently identically distributed as''.
This implies $Y_{ijk}$ are independent of each other and
$Y_{ijk} \sim N\left({\mu_i+\be_i \,X_{ij},\sigma_e^2} \right)$.
(c) The covariate and the treatment factors are independent.
Then the straight line fitted by ANCOVA to each treatment
can be written as $\widehat{Y}_{ij} \,=\,\widehat{\mu}_i+\bh_i\,X_{ij}$,
where $\widehat{Y}_{ij}$ is the predicted response for treatment $i$ at $X_{ij}$,
$\widehat{\mu}_i$ is the estimated intercept, and
$\bh_i $ is the estimated slope for treatment $i$.

In the analysis, these fitted lines can then be used to test
the following null hypotheses:
$$\text{(i) $H_o:\be_1=\be_2=\cdots=\be_t=0$ (All slopes are equal to zero).}$$
If $H_o $ is not rejected, then the covariate is not necessary in the model.
Then a regular one-way ANOVA can be performed
to test the equality of treatment means.
$$\text{(ii) $H_o:\be_1=\be_2=\cdots=\be_t $ (The slopes are equal).}$$
Depending on the conclusion reached here,
two types of models are possible for linear ANCOVA models -
parallel lines and nonparallel lines models.
If $H_o $ in (ii) is not rejected, then the lines are parallel,
otherwise they are nonparallel (\cite{milliken:2002}).
Throughout the article the terms
``parallel lines models (case)'' and ``equal slope models (case)'' will be
used interchangeably.
The same holds for ``nonparallel lines models (case)'' and
``unequal slopes models (case)''.

The parallel lines model is given by
\begin{equation}
\label{eqn:parallel-lines}
Y_{ijk} \,=\,\mu_i+\be \,X_{ij}+e_{ijk},
\end{equation}
where $\be $ is the common slope for all treatment levels.
With this model, testing the equality of the intercepts,
$H_o:\mu_1=\mu_2=\cdots =\mu_t $, is equivalent to
testing the equality of treatment means at any value of the covariate.
For the nonparallel lines case, the model is as in Equation \eqref{eqn:trt-spec-lines}
with at least one $\be_i$ being different for some $i=1,2,\ldots,t$.
So the comparison of treatments may give different results
at different values of the covariate.

\subsection{Analysis of Covariate-Adjusted Residuals}
\label{sec:Cov-Adj-Res}
First an overall regression line is fitted to the entire data set as:
\begin{equation}
\label{eqn:overall-reg-line}
\widehat{Y}_{ij}=\widehat{\mu}+\bh^\ast \,X_{ij},
\text{ for $i=1,2,\ldots,t$ and $j=1,2,\ldots,n_i$,}
\end{equation}
where $\widehat{\mu}$ is the estimated overall intercept and
$\bh^\ast$ is the estimated overall slope.
The residuals from this regression line are called \emph{covariate-adjusted residuals}
and are calculated as:
\begin{equation}
\label{eqn:residual-overall}
R_{ijk}=Y_{ijk} -\widehat{Y}_{ij}=Y_{ijk} -\widehat{\mu}-\bh^\ast \,X_{ij},
\text{for $i=1,2,\ldots,t$, $j=1,2,\ldots,n_i,$ and $k=1,2,\ldots,r_{ij}$,}
\end{equation}
where $R_{ijk}$ is the $k^{th}$ residual of treatment level $i$ at $X_{ij}$.

\subsubsection{ANOVA with or without HOV on Covariate-Adjusted Residuals}
\label{sec:ANOVA-Cov-Adj-Res}
In ANOVA with or without HOV procedures,
the covariate-adjusted residuals in Equation \eqref{eqn:residual-overall}
are taken to be the response values,
and tests of equal treatment means are performed on residual means.

The means model and assumptions for the one-way ANOVA with HOV on
these covariate-adjusted residuals are:
\begin{equation}
\label{eqn:model-on-residuals}
R_{ijk} \,=\,\rho_i+\ve_{ijk},
\text{ for $i=1,2,\ldots,t$, $j=1,2,\ldots,n_i,$ and $k=1,2,\ldots,r_{ij}$,}
\end{equation}
where $\rho_i$ is the mean residual for treatment $i$,
$\ve_{ijk}$ are the (independent) random errors such that
$\ve_{ijk} \sim N\left({0,\sigma_\ve^2} \right)$.
Notice  the common variance $\sigma_\ve^2 $ for all treatment levels.
However, $R_{ijk}$ are not independent of each other,
since $\sum_{i=1}^t {\sum_{j=1}^{n_i}{\sum_{k=1}^{r_{ij}}{R_{ijk}}}}=0$,
which also implies that the overall mean of the residuals is zero.
Hence the model in Equation \eqref{eqn:overall-reg-line} and
an effects model for residuals are identically parameterized.

For the nonparallel lines model in Equation \eqref{eqn:trt-spec-lines},
the residuals in Equation \eqref{eqn:residual-overall} will take the
form:
$$
R_{ijk}=
Y_{ijk} -\widehat{Y}_{ij}=
\mu_i+\beta_i \,X_{ij}+e_{ijk} -\left({\widehat{\mu}+\bh^\ast \,X_{ij}} \right)=
\left({\mu_i -\widehat{\mu}} \right)+\left({\beta_i -\bh^\ast}\right)\,X_{ij}+e_{ijk}.
$$
Hence, the influence of the covariate will be removed if and only if
\begin{equation}
\label{eqn:remove-influence-covariate}
\bh^\ast=\,\beta_i\;\;\;\mbox{for}\;\mbox{all}\;\;i=1,2,\ldots,t.
\end{equation}
Then taking covariate-adjusted residuals can only remove the influence of
the covariate when the treatment-specific lines in Equation \eqref{eqn:trt-spec-lines}
and the overall regression in Equation \eqref{eqn:overall-reg-line} are parallel.
Notice that the residuals from the treatment-specific models
in Equation \eqref{eqn:trt-spec-lines} cannot be used as response values in an ANOVA with HOV,
because treatment sums of squares of such residuals
are zero (\cite{ceyhan:masters-thesis}).

In ANOVA without HOV on covariate-adjusted residuals,
the only difference from ANOVA with HOV is that $\ve_{ijk}$ are the
(independent) random errors such that $\ve_{ijk} \sim N\left(0,\sigma_i^2 \right)$.
Notice the treatment-specific variance term $\sigma_i^2 $;
i.e., the homogeneity of the variances is not necessarily
assumed in this model.

Kruskal-Wallis (K-W) test is an extension of the
Mann-Whitney $U$ test to three or more groups;
and for two groups K-W test and Mann-Whitney $U$ test are equivalent (\cite{siegel:1988}).
K-W test on the covariate-adjusted residuals which are obtained
as in model \eqref{eqn:residual-overall}
tests the equality of the residual distributions
for all treatment levels.
Notice that contrary to the parametric models and tests in previous sections,
only the distributional equality is assumed, neither normality nor HOV.

\section{Comparison of the Methods}
\label{sec:comparison}
ANOVA with or without HOV or K-W test on covariate-adjusted residuals and
ANCOVA can be compared when the treatment-specific lines and
the overall regression line are parallel.
The null hypotheses tested by ``ANCOVA", ``ANOVA with or without HOV", and ``K-W test"
on covariate-adjusted residuals are
\begin{equation}
\label{eqn:null-ANCOVA-equal-intercept}
H_o:\mu_1=\mu_2=\cdots=\mu_t
\text{ (Intercepts are equal for all treatments.)}
\end{equation}

\begin{equation}
\label{eqn:null-ANOVA-cov-adj-residuals}
H_o:\rho_1=\rho_2=\cdots=\rho_t
\text{ (Residual means are equal for all treatments.)}
\end{equation}
and
\begin{equation}
\label{eqn:null-K-W-cov-adj-residuals}
H_o:F_{R_1}=F_{R_2}=\cdots=F_{R_t}
\text{ (Residuals have the same distribution for all treatments.),}
\end{equation}
respectively.

For more than two treatments the assumption of
parallelism is less likely to hold,
since only two lines with different slopes are sufficient to violate the condition.
With two treatments, the null hypotheses tested by
ANCOVA, ANOVA with or without HOV and K-W test on
covariate-adjusted residuals will be
\begin{equation}
\label{eqn:null-ANCOVA-equal-intercept-2class}
H_o:\mu_1=\mu_2
\text{ (or $\mu_1 -\mu_2=0$)}
\end{equation}

\begin{equation}
\label{eqn:null-ANOVA-cov-adj-residuals-2class}
H_o:\rho_1=\rho_2
\text{ (or $\rho_1 -\rho_2=0$)}
\end{equation}
and
\begin{equation}
\label{eqn:null-K-W-cov-adj-residuals-2class}
H_o:F_{R_1}=F_{R_2}
\text{ (or $R_1\stackrel{d}{=}R_2$)}
\end{equation}
respectively, where $\stackrel{d}{=}$ stands for ``equal in distribution''.

In Equation \eqref{eqn:null-ANOVA-cov-adj-residuals-2class},
$\rho_i$ can be estimated by the sample residual mean, $\overline {R}_{i..}$.
Combining the expressions in \eqref{eqn:residual-overall} and \eqref{eqn:model-on-residuals},
the residuals can be rewritten as
$$R_{ijk}=\rho_i+\ve_{ijk}=Y_{ijk} -\widehat{Y}_{ij}=
\left({\mu_i+\be_i \,X_{ij}+e_{ijk}}\right)-\left({\widehat{\mu}+\bh^\ast \,X_{ij}} \right),$$
$i=1,2$, $j=1,2,\ldots,n_i$, and $k=1,2,\ldots,r_{ij}$.
Averaging the residuals for treatment $i$ yields
\begin{equation}
\label{eqn:average-residuals}
\overline{R}_{i..}=\rho_i+\overline{\ve}_{i..}=
\mu_i+\be_i \,\overline{X}_{i.}+\overline{e}_{i..} -\widehat{\mu}-\bh^\ast\,\overline{X}_{i.},
\quad i=1,2
\end{equation}
where $\overline{X}_{i.}$ is the sample mean of covariate values for treatment $i$,
$\overline{e}_{i..}=\sum_{j=1}^{n_i}{{\sum_{k=1}^{r_{ij}}{e_{ijk}}} \mathord{\left/{\vphantom{{\sum_{k=1}^{r_{ij}}
{e_{ijk}}}{n_i}}} \right. \kern-\nulldelimiterspace}{n_i}}$ and
$\overline{\ve}_{i..}=\sum_{j=1}^{n_i} {{\sum_{k=1}^{r_{ij}}{\ve_{ijk}}} \mathord{\left/
{\vphantom{{\sum_{k=1}^{r_{ij}}{\ve_{ijk}}}{n_i}}}
\right. \kern-\nulldelimiterspace}{n_i}}$, $i=1,2$. Under the
assumptions of ANCOVA and ANOVA (with or without HOV) on covariate-adjusted
residuals, taking the expectations in \eqref{eqn:average-residuals} yields
\begin{equation}
\label{eqn:exp-average-residuals}
\E\left[{\overline{R}_{i..}} \right]=
\rho_i=\mu_i+\be_i \,\overline{X}_{i.} -\mu -\be^\ast \,\overline{X}_{i.}=
\mu_i -\mu+\left({\be_i \,-\be^\ast} \right)\,\overline{X}_{i.},
\quad
i=1,2,
\end{equation}
since $\E[\overline{e}_{i..}]=0$ and $\E[\overline{\ve}_{i..}]=0$, for $i=1,2$.
Hence $H_o$ in \eqref{eqn:null-ANOVA-cov-adj-residuals-2class} can be rewritten as
$H_o:\left({\mu_1 -\mu_2} \right)+\left({\be_1 -\be^\ast} \right)\overline
{X}_{1.} -\left({\be_2 -\be^\ast} \right)\overline{X}_{2.} \,=0$.
Then the hypotheses in Equations \eqref{eqn:null-ANCOVA-equal-intercept-2class}
and \eqref{eqn:null-ANOVA-cov-adj-residuals-2class} are equivalent iff
\begin{equation}
\label{eqn:hypo-equivalent}
\left({\be_1 -\be^\ast} \right)\overline{X}_{1.}=\left({\be_2-\be^\ast} \right)\overline{X}_{2.}
\end{equation}
Using condition \eqref{eqn:remove-influence-covariate} and repeating
the above argument for all pairs of treatments,
the condition in \eqref{eqn:hypo-equivalent} can be extended to more than two treatments.

Notice that the conditions that will imply \eqref{eqn:hypo-equivalent}
will also imply the equivalence of the hypotheses in
\eqref{eqn:null-ANCOVA-equal-intercept-2class} and \eqref{eqn:null-ANOVA-cov-adj-residuals-2class}.
The overall regression slope can be estimated as
\begin{equation}
\label{eqn:overall-reg-slope}
\bh^\ast=
\frac{\sum_{i=1}^2{\sum_{j=1}^{n_i}{\sum_{k=1}^{r_{ij}}
{\left({X_{ij} -\overline{X}_{..}}\right)
\left({Y_{ijk} -\overline{Y}_{...}} \right)}}}}{E_{xx}^\ast}
=\frac{\sum_{i=1}^2{\sum_{j=1}^{n_i}{\sum_{k=1}^{r_{ij}}
{\left({X_{ij} -\overline{X}_{..}} \right)}Y_{ijk}}}}{E_{xx}^\ast}
\end{equation}
where $\overline{X}_{..}$ is the overall covariate mean,
$\overline{Y}_{...}$ is the overall response mean, and
$$
E_{xx}^\ast=
\sum_{i=1}^2{\sum_{j=1}^{n_i}{r_{ij}\left({X_{ij} -\overline{X}_{..}} \right)^2}}=
\sum_{i=1}^2{\sum_{j=1}^{n_i}{r_{ij} \left({X_{ij} -\overline{X}_{..}}\right)X_{ij}}}.
$$
Furthermore the treatment-specific slope used in model \eqref{eqn:trt-spec-lines} is estimated as
$$
\bh_i=
\frac{\sum_{j=1}^{n_i}{\sum_{k=1}^{r_{ij}}{\left({X_{ij} -\overline{X}_{i.}} \right)}
\left({Y_{ijk} -\overline{Y}_{i..}} \right)}}{E_{xx,i}}
$$
where $E_{xx,i}=\sum_{j=1}^{n_i}{r_{ij} \left({X_{ij} -\overline{X}_{i.}} \right)^2}$,
and $\overline{Y}_{i..}$ is the mean response for treatment $i$.
Substituting $Y_{ijk} \,=\,\widehat{\mu}_i+\bh_i\,X_{ij}+{R}'_{ijk}$,
$i=1,2$, $j=1,2,\ldots,n_i$, and $k=1,2,\ldots,r_{ij}$ in Equation \eqref{eqn:overall-reg-slope}
where $\widehat{\mu}_i$ is the estimated intercept for treatment level $i$,
and ${R}'_{ijk}$ is the $k^{th}$ residual at $X_{ij}$
in model \eqref{eqn:trt-spec-lines},
the estimated overall slope becomes
$\displaystyle \bh^\ast=
\frac{\sum_{i=1}^2{\sum_{j=1}^{n_i}{\sum_{k=1}^{r_{ij}}{\left({X_{ij} -\overline{X}_{..}}\right)
\left({\widehat{\mu}_i+\bh_i \,X_{ij}+{R}'_{ijk}} \right)}}}}{E_{xx}^\ast}$.
With some rearrangements, we get
\begin{equation}
\label{eqn:est-overall-slope}
\begin{array}{l}
\bh^\ast=
\bh_i+\frac{\sum_{i=1}^2{\sum_{j=1}^{n_i}{r_{ij} \left({X_{ij} -\overline{X}_{..}}\right)\widehat{\mu}_i}}}{E_{xx}^\ast}+
\frac{\sum_{i=1}^2{\sum_{j=1}^{n_i}{\sum_{k=1}^{r_{ij}}{\left({X_{ij} -\overline{X}_{..}}
\right){R}'_{ijk}}}}}{E_{xx}^\ast} \\
\quad \;\,=\bh_i+\frac{\sum_{i=1}^2{\sum_{j=1}^{n_i}{r_{ij} \left({X_{ij} -\overline{X}_{..}}\right)\widehat{\mu}_i}}}{E_{xx}^\ast}+
\frac{\sum_{i=1}^2{\sum_{j=1}^{n_i}{\sum_{k=1}^{r_{ij}}{X_{ij}{R}'_{ijk}}}}}{E_{xx}^\ast}, \\
\end{array}
\end{equation}
since $\sum_{i=1}^2{\sum_{j=1}^{n_i}{\sum_{k=1}^{r_{ij}}{\overline{X}_{..}{R}'_{ijk}}}}=0$.
As $\E\left[{{R}'_{ijk}} \right]=0$,
taking the expectations of both sides of \eqref{eqn:est-overall-slope} yields
\begin{equation}
\label{eqn:beta-star}
\begin{array}{r}
\be^\ast=
\be_i+\frac{\mu_1 \left({\sum_{j=1}^{n_1}{r_{ij} \left({\overline{X}_{1j} -\overline{X}_{..}} \right)}} \right)+
\mu_2\left({\sum_{j=1}^{n_2}{r_{ij} \left({\overline{X}_{2j} -\overline{X}_{..}} \right)}} \right)}{E_{xx}^\ast} \\
=\be_i+\frac{\mu_1 n_1 \left({\overline{X}_{1.} -\overline{X}_{..}}
\right)+\mu_2 n_2 \left({\overline{X}_{2.} -\overline{X}_{..}}\right)}{E_{xx}^\ast} \\
\end{array}
\end{equation}
Under $H_o:\mu_1=\mu_2 $, \eqref{eqn:beta-star} reduces to $\be^\ast=\be_i$ iff
\begin{equation}
\label{eqn:beta-star=beta-i}
\frac{n_1 \left({\overline{X}_{1.} -\overline{X}_{..}} \right)+
n_2 \left({\overline{X}_{2.} -\overline{X}_{..}} \right)}{E_{xx}^\ast}=0
\end{equation}
provided that $E_{xx}^\ast \ne 0$.
Indeed, $E_{xx}^\ast=0$ will hold if and only if
all $X_{ij}$ are equal to a constant for each treatment $i$,
in which case, $\bh^\ast$ and $\bh_i$ will both be undefined.
The condition in \eqref{eqn:beta-star=beta-i} holds if
$\overline{X}_{1.}=\overline{X}_{2.}$ (=$\overline{X}_{..} )$.
Recall that $H_o:\rho_1=\rho_2 $ was shown to be equivalent
to $H_o:\mu_1=\mu_2 $ provided that
$\left({\be_1 -\be^\ast} \right)\overline{X}_{1.}=\left({\be_2 -\be^\ast} \right)\overline{X}_{2.}$,
which holds if $\overline{X}_{1.}=\overline{X}_{2.}$ and $\be_1=\be_2$.
So the null hypotheses in \eqref{eqn:null-ANCOVA-equal-intercept-2class}
and \eqref{eqn:null-ANOVA-cov-adj-residuals-2class} are equivalent when the
treatment-specific lines are parallel and treatment-specific means are equal
which implies the condition stated in \eqref{eqn:remove-influence-covariate}.

In general for $t$ treatments,
the hypotheses in \eqref{eqn:null-ANCOVA-equal-intercept} and
\eqref{eqn:null-ANOVA-cov-adj-residuals} can be tested
using an $F$ test statistics.
$H_{o}$ in \eqref{eqn:null-ANCOVA-equal-intercept} can be tested by
\begin{equation}
\label{eqn:F-ANCOVA}
F=\frac{MSTrt}{MSE},
\end{equation}
where $MSTrt$ is the mean square treatment for response values,
and $MSE$ is the mean square error for response values.
These mean square terms can be calculated as:
$$
MSTrt=
\frac{\sum_{i=1}^t{\sum_{j=1}^{n_i}{r_{ij}
\left[{\left({\overline{Y}_{i..} -\overline{Y}_{...}} \right)-
\bh_i \left({\overline{X}_{i.} -\overline{X}_{..}} \right)} \right]^2}}}{(t-1)}
$$
and
$$
MSE=
\frac{\sum_{i=1}^t{\sum_{j=1}^{n_i}{\sum_{k=1}^{r_{ij}}
{\left[{\left({Y_{ijk} -\overline{Y}_{i..}}\right)-
\bh_i \left({X_{ij} -\overline{X}_{i.}} \right)} \right]^2}}}}
{\left({n-(t+1)} \right)}.
$$
Note that $MSE$ has $\left({n-t-1}\right)$ degrees of freedom ({\it df})
since there are $(t+1)$ parameters ($\mu_i$ for $i=1,2,\ldots,t$ and $\be$) to estimate.
Therefore the test statistic in \eqref{eqn:F-ANCOVA} is distributed as
$F\sim F\left({t-1,n-t-1} \right)$.

Similarly, H$_{o}$ in \eqref{eqn:null-ANOVA-cov-adj-residuals} can be tested by
\begin{equation}
\label{eqn:F-ANOVA-on-Res}
F^\ast=\frac{MSTrt^\ast}{MSE^\ast},
\end{equation}
where $MSTrt^\ast$ is the mean square treatment for covariate-adjusted residuals,
and $MSE^\ast$ is the mean square error for covariate-adjusted residuals.
These mean square terms can be calculated as
$MSTrt^\ast=\frac{\sum_{i=1}^t{\sum_{j=1}^{n_i}{r_{ij}
\left({\overline{R}_{i..} -\overline{R}_{...}} \right)^2}}}{(t-1)}$ and
$MSE^\ast=\frac{\sum_{i=1}^t{\sum_{j=1}^{n_i}{\sum_{k=1}^{r_{ij}}
{\left({R_{ijk} -\overline{R}_{i..}} \right)^2}}}}{\left({n-\,t} \right)}$.
Using $\overline{R}_{i..}=\overline{Y}_{i..} -\widehat{\mu}-\bh^\ast \,\overline{X}_{i.}$,
$i=1,2,\ldots,t$, and
$\overline{R}_{...}=\overline{Y}_{...} -\widehat{\mu}-\bh^\ast \,\overline{X}_{..}$,
$$
MSTrt^\ast=
\frac{\sum_{i=1}^t{\sum_{j=1}^{n_i}{r_{ij}
\left[{\left({\overline{Y}_{i..} -\overline{Y}_{...}} \right)-
\bh^\ast\left({\overline{X}_{i.} -\overline{X}_{..}} \right)} \right]^2}}}{(t-1)}
$$
and
$$
MSE^\ast=
\frac{\sum_{i=1}^t{\sum_{j=1}^{n_i}{\sum_{k=1}^{r_{ij}}
{\left[{\left({Y_{ijk} -\overline{Y}_{i..}}\right)-
\bh^\ast \left({X_{ij} -\overline{X}_{i.}} \right)}\right]^2}}}}{\left({n-\,t} \right)}.
$$
It might seem that $MSE^\ast$ has $\left({n-\,t} \right)$ degrees of freedom ({\it df}),
since there are $t$ parameters ($\rho_i$ for $i=1,2,\ldots,t$) to estimate,
so the test statistic in Equation \eqref{eqn:F-ANOVA-on-Res} is
distributed as $F^\ast \sim F\left({t-1,n-t}\right)$.
However, there is one more restriction in test \eqref{eqn:null-ANOVA-cov-adj-residuals-2class}.
Since $\sum_{i=1}^2{\sum_{j=1}^{n_i}{\sum_{k=1}^{r_{ij}}{R_{ijk}}}}=0$,
then $F^\ast$ should actually be distributed as $F^\ast \sim F\left({t-1,n-t-1} \right)$.
\cite{atchley:1976} did not suggest this adjustment in {\it df}, and
\cite{beaupre:1998} used the method without such an adjustment.
That is, in both sources $F\left({t-1,n-t}\right)$ is used for inference.
So, in this article {\it df} for $MSE^\ast$ has been set at
$\left({n-\,t} \right)$ as in literature.

Notice that, the $F$-statistics in \eqref{eqn:F-ANCOVA} and \eqref{eqn:F-ANOVA-on-Res} become
\begin{equation}
\label{eqn:F-ANCOVA-long}
F=
\frac{{\sum_{i=1}^t{\sum_{j=1}^{n_i}{r_{ij} \left[
{\left({\overline{Y}_{i..} -\overline{Y}_{...}} \right)-
\bh\left({\overline{X}_{i.} -\overline{X}_{..}} \right)} \right]^2}}}
\mathord{\left/{\vphantom
{{\sum_{i=1}^t{\sum_{j=1}^{n_i}{r_{ij} \left[{\left(
{\overline{Y}_{i..} -\overline{Y}_{...}} \right)-\bh\left({\overline{X}_{i.}
-\overline{X}_{..}} \right)} \right]^2}}}{(t-1)}}} \right.
\kern-\nulldelimiterspace}{(t-1)}}{{\sum_{i=1}^t
{\sum_{j=1}^{n_i}{\sum_{k=1}^{r_{ij}}{\left[{\left(
{Y_{ijk} -\overline{Y}_{i..}} \right)-\bh\left({X_{ij} -\overline{X}_{i.}} \right)} \right]^2}}}}
\mathord{\left/{\vphantom
{{\sum_{i=1}^t{\sum_{j=1}^{n_i}
{\sum_{k=1}^{r_{ij}}{\left[{\left({Y_{ijk} -\overline{Y}_{i..}}
\right)-\bh\left({X_{ij} -\overline{X}_{i.}} \right)} \right]^2}}}
}{\left({n-\,(t+1)} \right)}}} \right. \kern-\nulldelimiterspace}{\left({n-\,(t+1)} \right)}},
\end{equation}
and
\begin{equation}
\label{eqn:F-ANOVA-on-Res-long}
F^\ast=
\frac{{\sum_{i=1}^t{\sum_{j=1}^{n_i}{r_{ij}
\left[{\left({\overline{Y}_{i..} -\overline{Y}_{...}} \right)-
\bh^\ast \left({\overline{X}_{i.} -\overline{X}_{..}} \right)} \right]^2}}}
\mathord{\left/{\vphantom{{\sum_{i=1}^t{\sum_{j=1}^{n_i
}{r_{ij} \left[{\left({\overline{Y}_{i..} -\overline{Y}_{...}} \right)-
\widehat{\be}^\ast \left({\overline{X}_{i.} -\overline{X}_{..}} \right)} \right]^2}}}
{(t-1)}}} \right. \kern-\nulldelimiterspace}{(t-1)}}{{\sum_{i=1}^t
{\sum_{j=1}^{n_i}{\sum_{k=1}^{r_{ij}}{\left[{\left({Y_{ijk} -
\overline{Y}_{i..}} \right)-\bh^\ast \left({X_{ij} -\overline{X}_{i.}} \right)} \right]^2}}}}
\mathord{\left/{\vphantom {{\sum_{i=1}^t{\sum_{j=1}^{n_i}
{\sum_{k=1}^{r_{ij}}{\left[{\left({Y_{ijk} -\overline{Y}_{i..}}
\right)-\bh^\ast \left({X_{ij} -\overline{X}_{i.}} \right)}\right]^2}}}}
{\left({n-\,t} \right)}}} \right.
\kern-\nulldelimiterspace}{\left({n-\,t} \right)}},
\end{equation}
respectively.
For two treatments, $t=2$ will be used in Equations \eqref{eqn:F-ANCOVA-long} and \eqref{eqn:F-ANOVA-on-Res-long},
then the test statistics will be distributed as
$F\sim F\left({1,n-3} \right)$ and $F^\ast \sim F\left({1,n-2} \right)$,
and they can be used to test the hypotheses in Equations
\eqref{eqn:null-ANCOVA-equal-intercept-2class} and \eqref{eqn:null-ANOVA-cov-adj-residuals-2class},
respectively.
Furthermore, with two treatments, note that
$F\,\stackrel{d}{=}\,\mathscr T^2\left({n-3} \right)$ and $F^\ast \,\stackrel{d}{=}\,\mathscr T^2\left( {n-2}\right)$
and $\mathscr T\left( n \right)$ is the $t$-distribution with $n$ $df$.
As $n \to \infty $,
both $F$ and $F^\ast$ will converge in distribution to $\chi_1^2$.
So $F$ and $F^\ast$ will have
similar observed significance levels (i.e., $p$-values) and
similar scores for large $n$.
Similar decisions for testing \eqref{eqn:null-ANCOVA-equal-intercept-2class} and
\eqref{eqn:null-ANOVA-cov-adj-residuals-2class} will be
reached if the calculated test statistics are similar;
i.e.,
$F\approx F^\ast$ for large $n$.
Likewise,
$F$ in \eqref{eqn:F-ANCOVA} and $F^\ast$ in \eqref{eqn:F-ANOVA-on-Res} will have similar
distributions for large $n$.

For the case of two treatments, comparing $F$ and $F^\ast$, it can be seen
that $F$ and $F^\ast$ are similar if $\bh^\ast \approx \widehat {\be}_i$ for large $n$.
The same argument holds for the test statistics
in the general case of more than two treatments for large $n$.
The test statistics will lead to similar decisions,
if $\bh^\ast \approx \bh_i$ as $n$ increases.
That is, the overall regression line fitted to the entire data set
should be approximately parallel to the fitted
treatment-specific regression lines for the test statistics
$F$ and $F^\ast$ to be similar.
If $\bh^\ast \approx \bh_i$,
then ANOVA with or without HOV on covariate-adjusted residuals and
ANCOVA will give similar results.
Consequently, it is expected that the ANCOVA and
ANOVA with HOV or without HOV methods give similar results as treatment-specific
covariate means gets closer for the parallel lines case.

The above discussion is based on normality of error terms with HOV.
Without HOV the {\it df} of the $F$-tests are calculated with
Satterthwaite approximation (\cite{kutner:2004}).
On the other hand, K-W test does require neither normality nor HOV,
but implies a more general hypothesis
$H_o:F_{R_1}=F_{R_2}$, in the sense that $H_o $ would imply
$H_o:\rho_1=\rho_2$ without the normality assumption.
However, the null hypothesis in
Equation \eqref{eqn:null-ANOVA-cov-adj-residuals-2class} implicitly assumes normality.

\section{Monte Carlo Simulation Analysis}
\label{sec:MonteCarlo}
Throughout the simulation only two treatments ($t=2)$ are used for the comparison of methods.
In the simulation, sixteen different cases are considered for comparison
(see Table \ref{tab:list-sim-cases}).

\subsection{Sample Generation for Null and Alternative Models}
\label{sec:Sample-Gen-Null-Alt}
Without loss of generality,
the slope in model \eqref{eqn:parallel-lines} is arbitrarily
taken to be 2 and the intercept is chosen to be 1.
So the response values for the treatments are generated as
\begin{equation}
\label{eqn:generate-resp-trt1}
\text{(i) $Y_{1jk}=1+2X_{1j}+e_{1jk}$, $j=1,2,\ldots,n_1$ and $k=1,2,\ldots,r_{1j}$ for treatment 1}
\end{equation}
with $e_{1jk} \stackrel{iid}{\sim}F_1 $, where $F_1 $ is the error distribution for treatment 1.

\begin{equation}
\label{eqn:generate-resp-trt2}
\text{(ii) $Y_{2jk}=(1+0.02q)+2X_{2j}+e_{2jk}$, $j=1,2,\ldots,n_2$ and $k=1,2,\ldots,r_{2j}$ for treatment 2}
\end{equation}
with $e_{2jk} \stackrel{iid}{\sim}F_2 $, where $F_2 $ is the error distribution for treatment 2 and
$q$ is introduced to obtain separation between the parallel lines.
In \eqref{eqn:generate-resp-trt1} and \eqref{eqn:generate-resp-trt2},
$X_{ij}$ is the $j^{th}$ generated value of the covariate in treatment $i$,
$Y_{ijk}$ is the response value for treatment level $i$ at $X_{ij}$ for $i=1,2$,
$e_{ijk}$ is the $k^{th}$ random error term.
The covariate ranges, sample sizes ($n_1$ and $n_2$),
error distributions ($F_1$ and $F_2$) for the two treatments,
and the number of replicates (reps) at each value of $X_{ij}$ are summarized in Table \ref{tab:list-sim-cases}.
In the context of model \eqref{eqn:parallel-lines} the common slope is
$\be=2$, and $\mu_1=1$ and $\mu_2=(1+0.02\,q)$ are
the intercepts for treatment levels 1 and 2, respectively.

Then as $q$ increases the treatment-specific response
means become farther apart at each covariate value and
the power of the tests is expected to increase.
The choice of 0.02 for the increments is based on time
and efficiency of the simulation process.
$q$ is incremented from 1 to $m_u$ in case-$u$,
for $u=1,2,...,\;16$ (Table \ref{tab:list-sim-cases})
where $m_u $ is estimated by the standard errors of the intercepts of the
treatment-specific regression lines.
In the simulation no further values of $q$ are chosen
when the power is expected to approach 1.00 that occurs
when the intercepts are approximately 2.5 standard errors apart,
as determined by equating the intercept difference,
$0.02\,q=2.5\,s_{\widehat{\mu}_i}$, with $q$ replaced by $m_u$.
A pilot sample of size 6000 is generated
($q=0,1,2,3,4,5$ with 1000 samples at each $q$),
and maximum of the standard errors of the intercepts is recorded.
Then $m_u \cong{2.5\max_i (s_{\widehat {\mu}_i} )}
\mathord{\left/{\vphantom{{2.5\max_i (s_{\widehat{\mu}_i} )}{0.02}}} \right. \kern-\nulldelimiterspace}{0.02}$
for $i=1,2$ in case $u$.

All cases labeled with ``a'' have one replicate and
all cases labeled with ``b'' have two replicates per covariate value, henceforth.
For example, in case 1a the most general case is simulated with $iid\;N(0,1)$ error variances,
and 20 uniformly randomly generated covariate values
in the interval $(0,10)$ for both treatments.
In case 1b, the data is generated as in case 1a with two replicates per covariate value.

In cases 1, 5-8, 9, and 12-16, error variances are homogeneous;
in cases 1, and 5-8 error terms are generated as $iid\;N(0,1)$.
In case 9, error terms are generated as {\it iid} $\U\left({-\sqrt 3,\sqrt 3}\right)$;
in case 12, error terms are {\it iid} $DW\left({0,1,3} \right)$,
double-Weibull distribution with location parameter 0,
scale parameter 1, and shape parameter 3 whose pdf is
$\displaystyle f(x)=\frac{3}{2}x^2\exp \left(-\vert x\vert^3\right)$ for all $x$;
in case 13, error terms are {\it iid}
$\sqrt{48} \left({\be \left({6,2} \right)-3 \mathord{\left/ {\vphantom{3 4}} \right.
\kern-\nulldelimiterspace} 4} \right)$
where $\be \left({6,2}\right)$ is the Beta distribution
with shape parameters 6 and 2 whose pdf is $\displaystyle f(x)=42x^5(1-x)\I(0<x<1)$
where $\I(\cdot)$ is the indicator function;
in case 14, error terms are {\it iid} $\chi_2^2 -2$ where $\chi_2^2 $ is the chi-square
distribution with 2 {\it df};
in case 15, error terms are {\it iid} $LN(0,1)-e^{1 \mathord{\left/{\vphantom{1 2}} \right. \kern-\nulldelimiterspace} 2}$
where $LN(0,1)$ is the log-normal distribution
with location parameter 0 and scale parameter 1 whose pdf is
$\displaystyle f(x)=\frac{1}{x\sqrt{2\pi}}\exp \left({-\frac{1}{2}(\log x)^2} \right)\,\I(x>0)$,
and in case 16, error terms are {\it iid} $N(0,2)$ for treatment 1 and
iid $\chi_2^2 -2$ for treatment 2.

In cases 2-4 heterogeneity of variances for normal error terms is
introduced either by unequal but constant variances (case 2),
unequal but a combination of constant and $x$-dependent variances (case 3),
or equal and $x$-dependent variances (case 4).
In case 10 error terms are {\it iid} $\U\left({-\sqrt 3,\sqrt 3} \right)$ for treatment 1
and {\it iid} $\U\left({-2\sqrt 3,2\sqrt 3} \right)$ for treatment 2;
in case 11, error terms are {\it iid} $\U\left({-\sqrt3,\sqrt 3} \right)$ treatment 1
and {\it iid} $\U\left({-\sqrt{x},\sqrt{x}} \right)$ for treatment 2.

The choice of constant variances is arbitrary,
but the error term distributions for constant variance cases are picked
so that their variances are roughly between 1 and 6.
However, $x$-dependence of variances is a realistic but not a general case,
since any function of $x$ could have been used.
For example, \cite{beaupre:1998} who explored the differences in metabolism
(O$_{2}$ consumption) of the Western diamondback rattlesnakes with respect to their sex,
the O$_{2}$ consumption was measured for males,
non-reproductive females, and vitellogenic females.
To remove the influence of body mass which was deemed as a covariate on O$_{2}$ consumption,
ANOVA with HOV on covariate-adjusted residuals was performed.
In their study, the variances of O$_{2}$ consumption for sexual groups
have a positive correlation with body mass.
In this study, $\sqrt{x}$ is taken as the variance term to simulate such a case.
Heterogeneity of variances conditions violate one of the assumptions for
ANCOVA and ANOVA with HOV on covariate-adjusted residuals,
and are simulated in order to evaluate the sensitivity of the methods to such violations.
The unequal variances in cases 2 and 3 were arbitrarily assigned to the treatments
since all the other restrictions are the same for treatments at each of these cases.
In case 5, different sample sizes are taken from that of other cases to see the
influence of unequal sample sizes.

In cases 1-8, error terms are generated from a normal distribution.
In cases 9-15, non-normal distributions for error generation are employed.
In cases 9-12, the distribution of the error variances are symmetric around 0,
while in cases 13-15 the distributions of the error terms are not symmetric around 0.
Notice that cases 13-15 are normalized to have zero mean,
and furthermore case 13 is scaled to have unit variance.
The influence of non-normality and asymmetry of the distributions
are investigated in these cases.
In case 16, the influence of distributional differences
(normal vs asymmetric non-normal) in the error term is investigated.

\begin{figure}[htbp]
\centerline{\includegraphics[width=5.08in,height=2.97in]{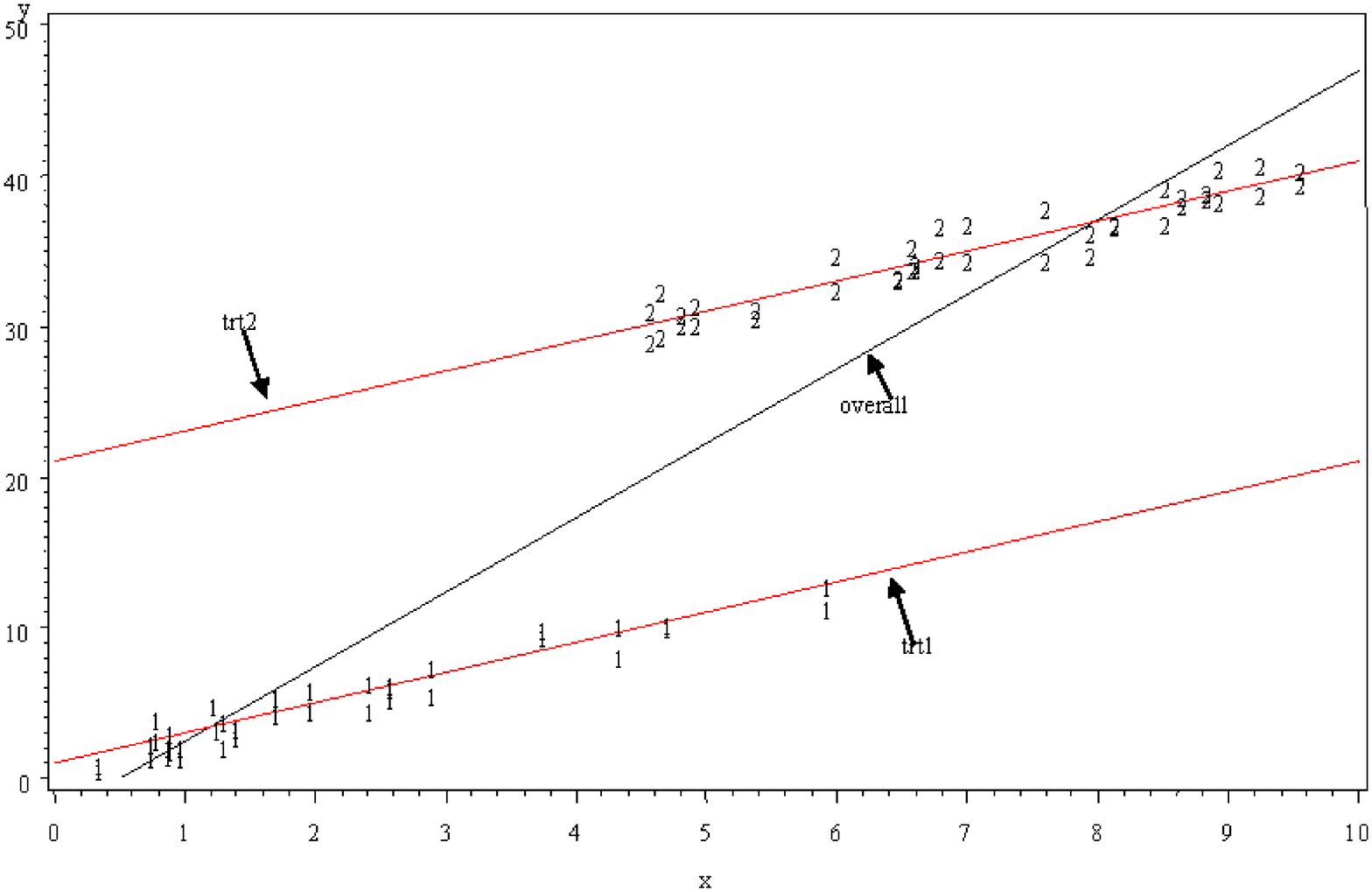}}
\caption{
\label{fig:case6}
A sample plot for case 6,
where observations from treatment $i$ are marked with $i$, for $i=1,2$,
trt $i$= fitted regression line for treatment $i$, $i=1,2$;
overall= overall fitted regression line.
}
\end{figure}

\begin{figure}[htbp]
\centerline{\includegraphics[width=5.12in,height=2.96in]{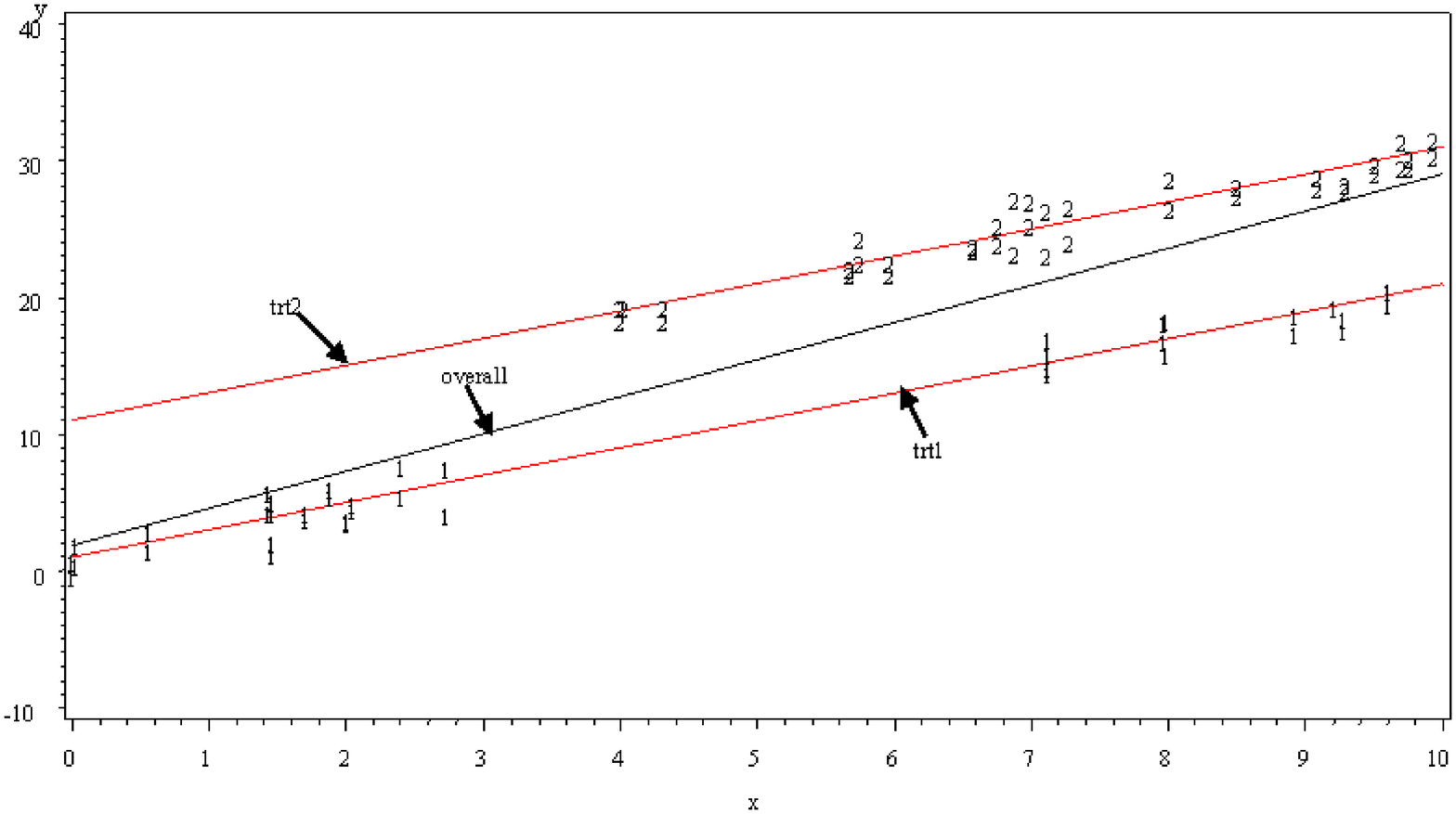}}
\caption{
\label{fig:case7}
A sample plot for case 7. Labeling is as in Figure \ref{fig:case6}.}
\end{figure}

In cases 1-5 and 9-16, covariates are uniformly randomly generated,
without loss of generality, in $(0,10)$,
hence $\overline{X}_{1.} \approx \overline{X}_{2.}$ is expected to hold.
In these cases the influence of replications (or magnitude of equal sample sizes),
heterogeneity of variances, and non-normality of the variances on the methods are investigated.
Cases 6-8 address the issue of clustering which might result naturally in a data set.
Clustering occurs if the treatments have distinct or
partially overlapping ranges of covariates.
Extrapolation occurs if the clusters are distinct
or the mean of the covariate is not within the covariate clusters
for at least one treatment.
In case 6 there is a mild overlap of the covariate clusters for treatments 1 and 2,
such that covariates are uniformly randomly generated within $(0,6)$ for treatment 1,
and $(4,10)$ for treatment 2,
so $\overline{X}_{1.}$ and $\overline{X}_{2.}$ are expected to be different.
In fact, this case is expected to contain the largest difference
between $\overline{X}_{1.}$ and $\overline{X}_{2.}$.
See Figure \ref{fig:case6} for a realization of case 6.
In case 7 treatment 1 has two clusters,
such that each treatment 1 covariate is randomly assigned to either $(0,3)$ or $(7,10)$ first,
then the covariate is uniformly randomly generated in that interval.
Treatment 2 covariates are generated uniformly within the interval of $(4,10)$.
Note that $\overline{X}_{1.}$ and $\overline{X}_{2.}$ are expected to be very different,
but not as much as case 6.
See Figure \ref{fig:case7} for a realization of case 7.
Notice that the second cluster of treatment 1 is completely inside the
covariate range of treatment 2.
These choices of clusters are inspired by the research of \cite{beaupre:1998}
which dealt with O$_{2}$ consumption of rattlesnakes.
In case 8 treatment 1 has two clusters,
each treatment 1 covariate is uniformly randomly generated
in the randomly selected interval of either $(0,4)$ or $(6,10)$.
Treatment 2 covariates are uniformly randomly generated in the interval $(3,7)$.
Hence $\overline{X}_{1.}$ and $\overline{X}_{2.}$ are expected to be similar.
Notice that treatment 2 cluster is in the middle
of the treatment 1 clusters with mild overlaps.

\subsection{Monte Carlo Simulation Results}
\label{sec:Monte-Carlo-Results}
In this section, the empirical size and power comparisons for the methods
discussed are presented.

\subsubsection{Empirical Size Comparisons}
\label{sec:empirical-size}
In the simulation process,
to estimate the empirical sizes of the methods in question,
for each case enumerated in Table \ref{tab:list-sim-cases},
$N_{mc}=10000$ samples are generated with $q=0$
using the relationships in \eqref{eqn:generate-resp-trt1} and \eqref{eqn:generate-resp-trt2}.
Out of these 10000 samples the number of significant treatment differences
detected by the methods is recorded.
The number of differences detected concurrently by each pair of methods is also recorded.
The nominal significance level used in all these tests is $\al=0.05$.
Based on these detected differences,
empirical sizes are calculated as
$\ah_i={\nu_i} \mathord{\left/{\vphantom{{\nu_i}{N_{mc}}}} \right.\kern-\nulldelimiterspace}{N_i}$
where $\nu_i$ are number of significant treatment differences detected by method $i$
with method 1 being ANCOVA,
method 2 being ANOVA with HOV,
method 3 being to ANOVA without HOV,
and method 4 being K-W test on covariate-adjusted residuals.
Furthermore the proportion of differences detected concurrently
by each pair of methods is
$\ah_{i,j}={v_{i,j}} \mathord{\left/{\vphantom{{v_{i,j}}{N_{mc}}}}
\right. \kern-\nulldelimiterspace}{N_{mc}}$,
where $N_{mc}=10000$ and $\nu_{i,j}$ is the number of significant treatment differences
detected by methods $i$,$j$, with $i \ne j.$
For large $N_{mc}$, $\ah_i \dot{\sim}N(\al_i,\sigma_{\al_i}^2 )$, $i=1,2,3,4$,
where $\dot{\sim}$ stands for ``approximately distributed as'',
$\al_i$ is the proportion of treatment differences,
$\sigma_{\al_i}^2={\al_i (1-\al_i)} \mathord{\left/{\vphantom{{\al_i (1-\al_i )}{N_{mc}}}}
\right. \kern-\nulldelimiterspace}{N_{mc}}$
is the variance of the unknown proportion,
$\al_i$ whose estimate is $\ah_i$.
Using the asymptotic normality of proportions for large $N_{mc}$,
the 95{\%} confidence intervals are constructed for empirical sizes of the methods
(not presented) to see whether they contain the nominal significance level,
0.05 and the 95{\%} confidence interval for the difference in the proportions
(not presented either) to check whether the sizes are significantly different
from each other.

\begin{figure}[htbp]
\centerline{\includegraphics[width=3.83in,height=2.99in]{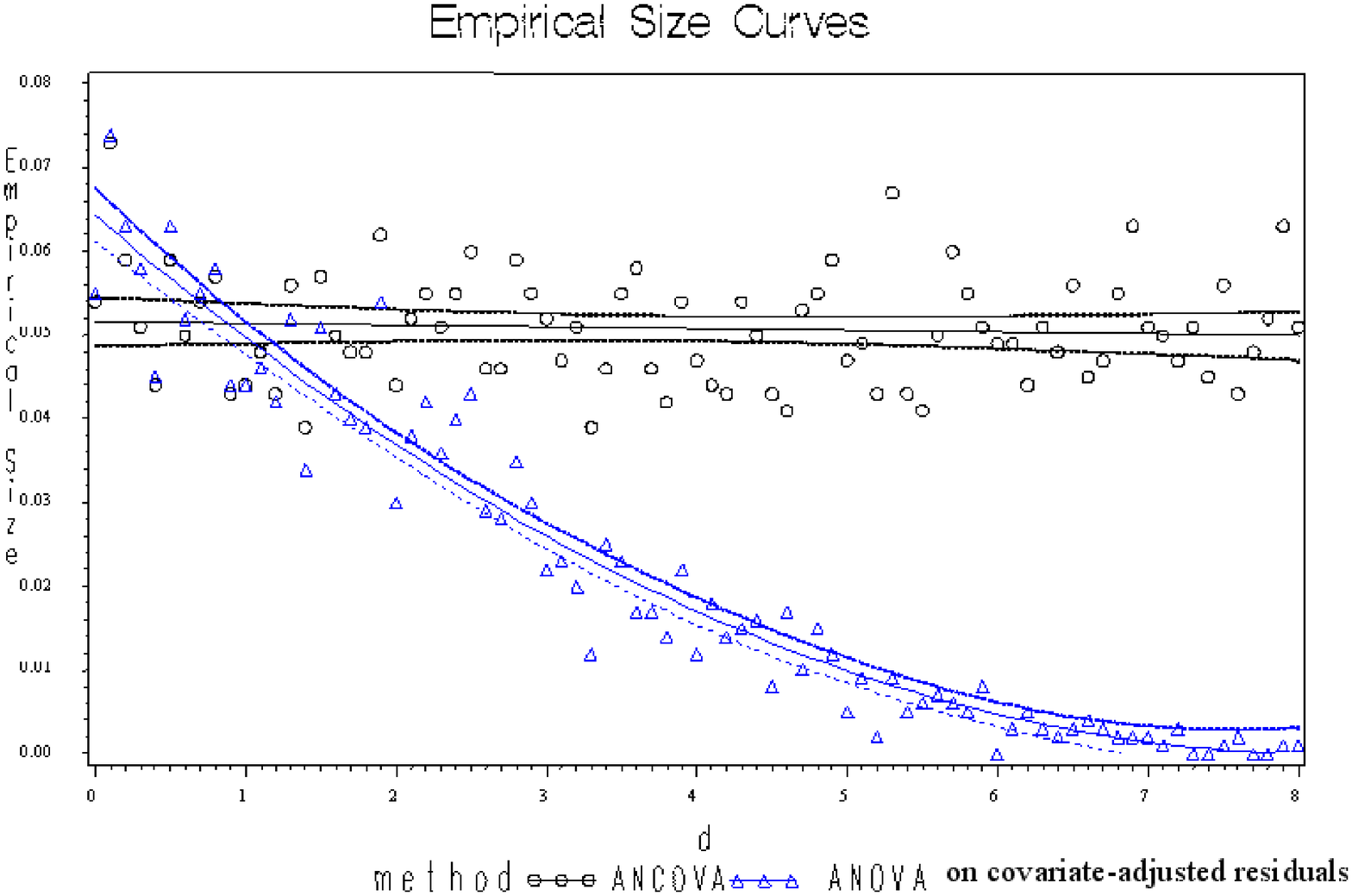}}
\caption{
\label{fig:emp-size}
Empirical sizes for ANCOVA and ANOVA on covariate-adjusted residuals
versus the distance between the treatment-specific means,
$d=\overline{X}_{1.} -\overline{X}_{2.}$, with the corresponding 95{\%} confidence bands.
}
\end{figure}

The empirical size estimates in cases 1a-16a and 1b-2b are presented in Table \ref{tab:emp-size}.
Observe that ANCOVA method is liberal in case 2a and
conservative at cases 14a and 15a, and
has the desired nominal level 0.05 for the other cases.
The liberalness in case 2a weakens as the number of replicates is doubled (see case 2b).
ANOVA with or without HOV are liberal in cases 1a, 2a, and 3a,
and conservative in cases 6a-8a, and 14a-15a and
have the desired nominal level for the other cases.
However, the liberalness of the tests weakens in cases 1a-3a,
as the number of replicates is doubled (see cases 1b-3b).
K-W test is liberal in cases 1a-3a, 10a, 11a and 16a,
and conservative in cases 6a, 7a, and 14a,
and has the desired nominal level for the other cases.
Liberalness of the test in case 1a weakens as the number of replicates is doubled (see case 1b).
Notice that the ANCOVA method has the desired size
when the error term is normally distributed or has a symmetric distribution,
tends to be slightly liberal when HOV is violated,
and is conservative when error distribution is non-normal and not symmetric.
On the other hand,
ANOVA with or without HOV have about the same size for all cases.
Both methods have the desired size when error terms are normally distributed,
or have symmetric distribution,
and the covariates have similar means.
When error terms are normal without HOV,
both methods are liberal with ANOVA without HOV being less liberal.
When error terms are non-normal with asymmetric distributions,
both methods tend to be slightly conservative.
But, when the covariate means are extremely different,
both methods are extremely conservative (see cases 6 and 7).
See Figure \ref{fig:emp-size} for the empirical size estimates
for ANCOVA and ANOVA with HOV on covariate-adjusted
residuals as a function of distance between treatment-specific means.
As the distance between treatment-specific means increase the empirical size for
the ANOVA with HOV on covariate-adjusted residuals decreases,
while the empirical size for ANCOVA is stable about the desired nominal level 0.05.
K-W test has the desired level when error terms have symmetric and identical distributions,
is liberal when errors have the same distribution without HOV and different distributions,
and is conservative when errors have asymmetric distributions
provided the covariates have similar means.
But when the covariate means are very different,
KW test is also extremely conservative (see cases 6 and 7).

Moreover, observe that when the covariates have similar means,
ANCOVA and ANOVA (with or without HOV) methods have similar empirical sizes.
These three methods have similar sizes as K-W test when the error distributions have HOV.
Without HOV, K-W test has significantly larger empirical size.
When the covariate means are considerably different,
ANCOVA method has significantly larger size than others.
ANOVA with or without HOV methods have similar empirical sizes for all cases.

As seen in Table \ref{tab:prop-agreement},
the proportion of agreement between the empirical size
estimates are usually not significantly different from the minimum of each
pair of tests for ANCOVA and ANOVA with or without HOV,
but the proportion of agreement is usually significantly smaller for
the cases in which K-W test is compared with others.
Therefore,
when covariate means are similar,
ANCOVA and ANOVA with or without HOV have the same null hypothesis,
with similar acceptance/rejection regions,
while K-W test has a different null hypothesis hence different acceptance/rejection regions.
When covariate means are different,
ANCOVA and ANOVA methods have different acceptance/rejection regions,
and K-W test has a different null hypothesis.
Both ANOVA methods have the same null hypothesis,
and have similar acceptance/rejection regions for this simulation study.

\subsubsection{Empirical Power Comparisons}
\label{sec:empirical-power}
\begin{figure}[htbp]
\centering
\rotatebox{-90}{ \resizebox{3. in}{!}{ \includegraphics{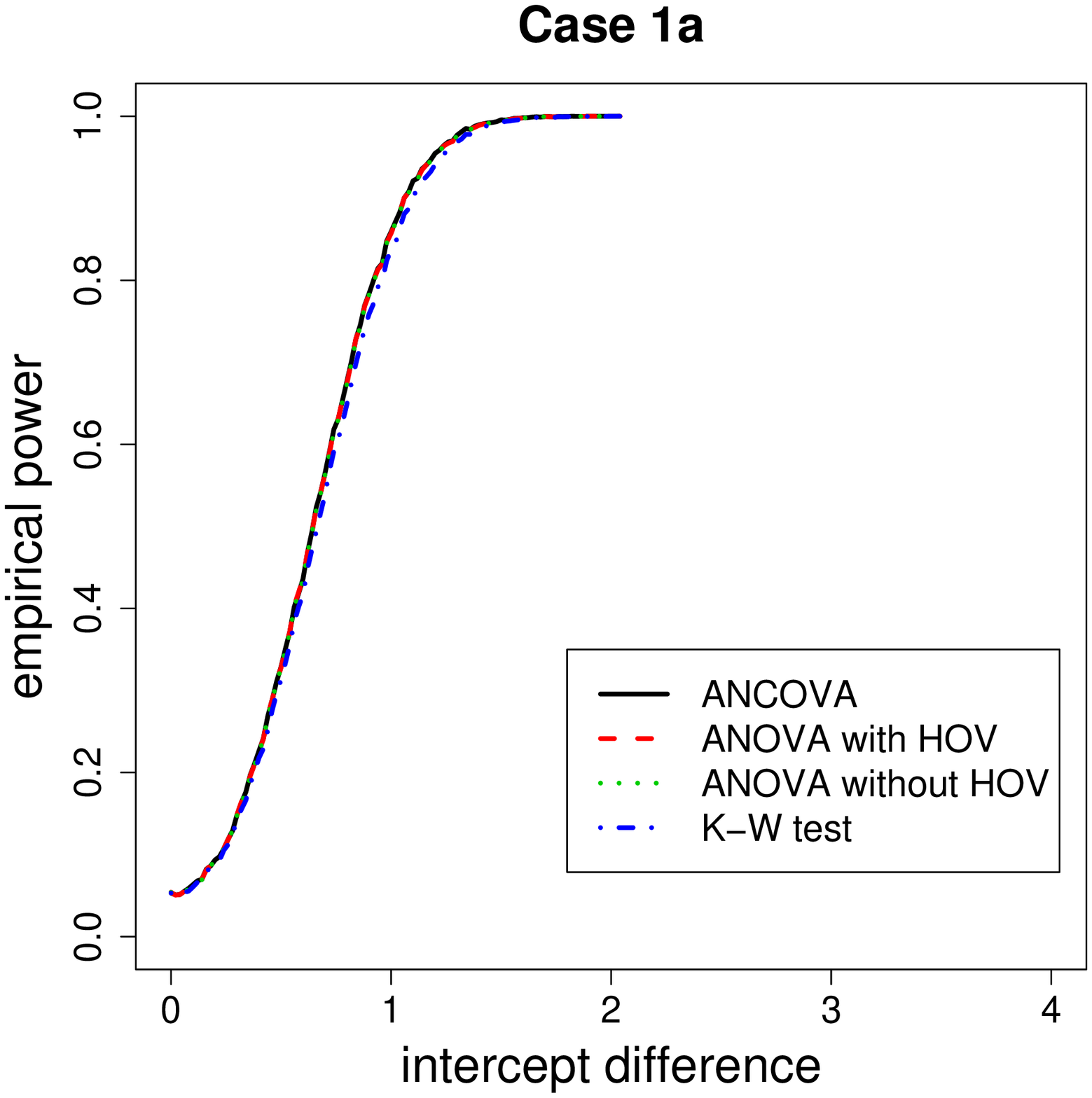}}}
\rotatebox{-90}{ \resizebox{3. in}{!}{ \includegraphics{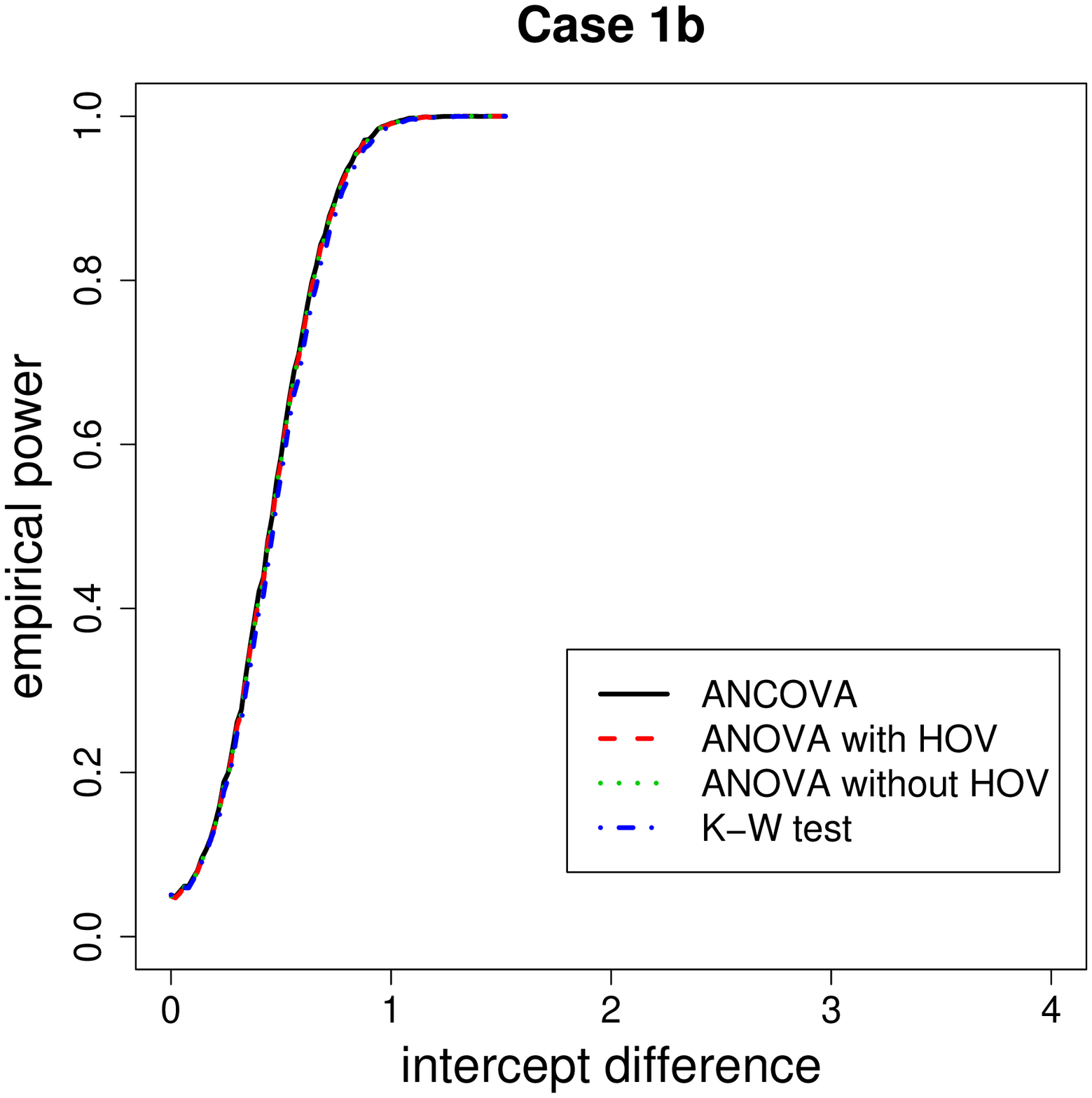}}}
\caption{
\label{fig:emp-power1-2a-b}
Empirical power estimates versus intercept difference for cases 1a and 1b.}
\end{figure}

\begin{figure}[htbp]
\centering
\rotatebox{-90}{ \resizebox{3. in}{!}{ \includegraphics{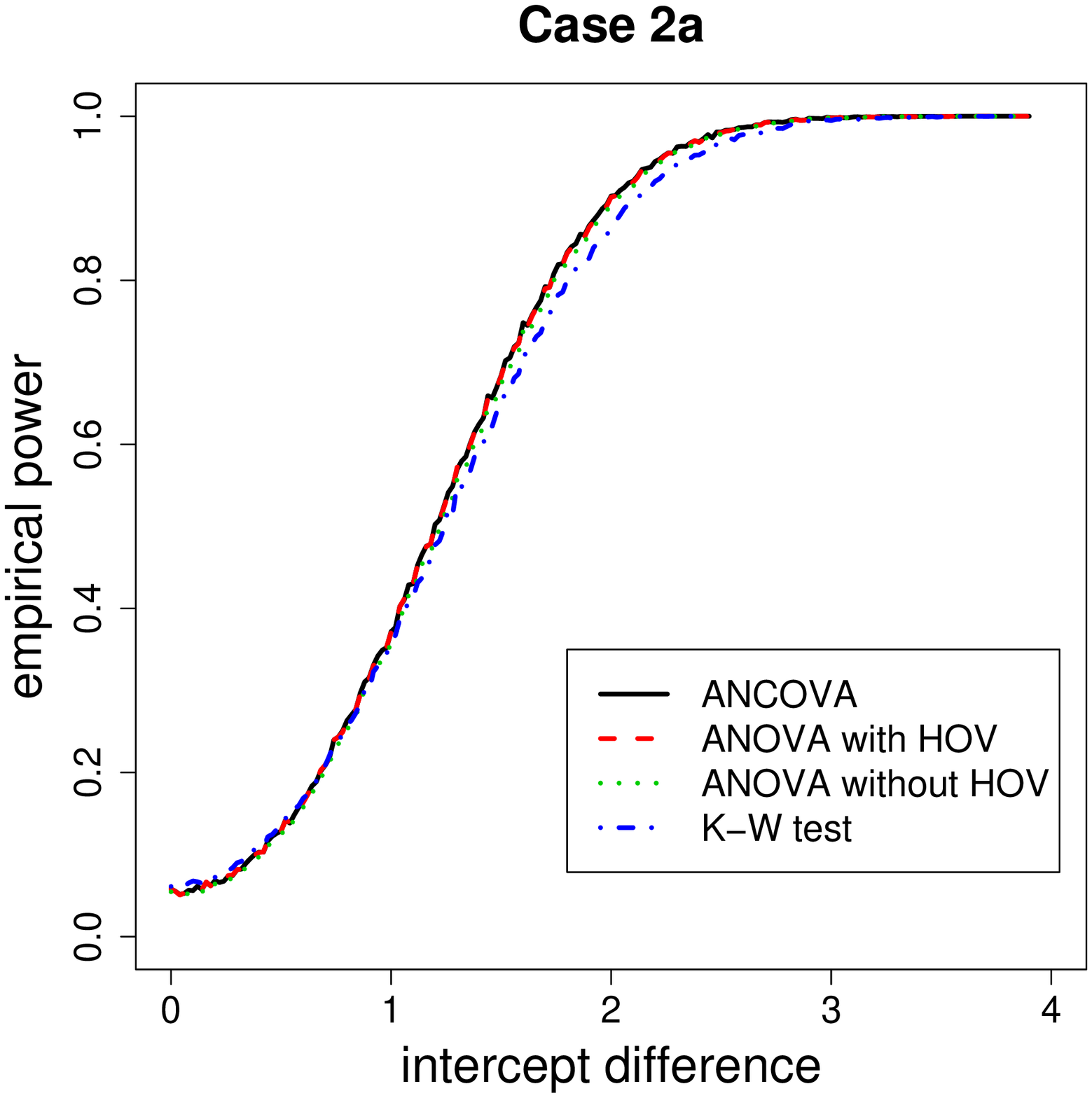}}}
\rotatebox{-90}{ \resizebox{3. in}{!}{ \includegraphics{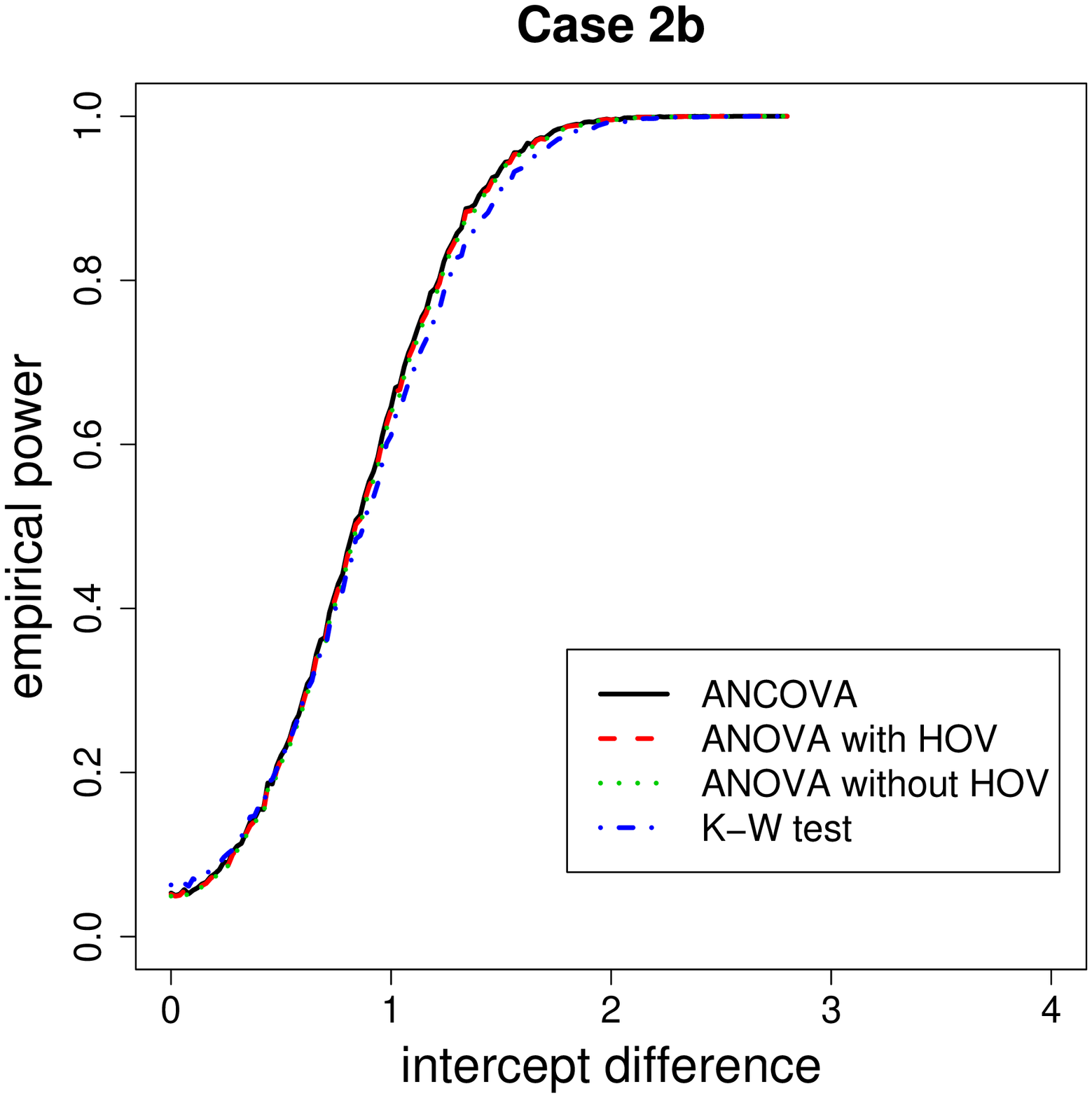}}}
\rotatebox{-90}{ \resizebox{3. in}{!}{ \includegraphics{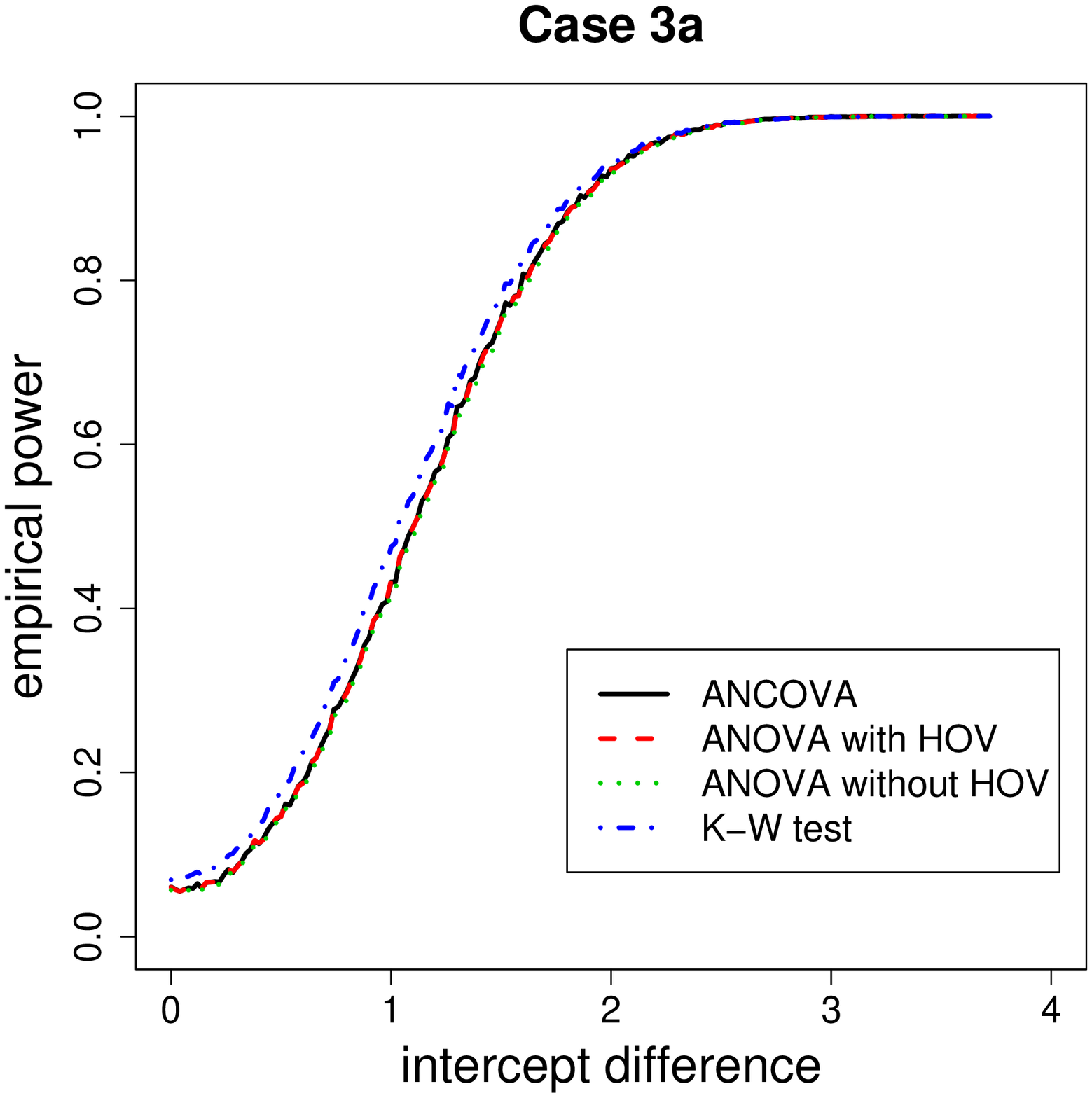}}}
\rotatebox{-90}{ \resizebox{3. in}{!}{ \includegraphics{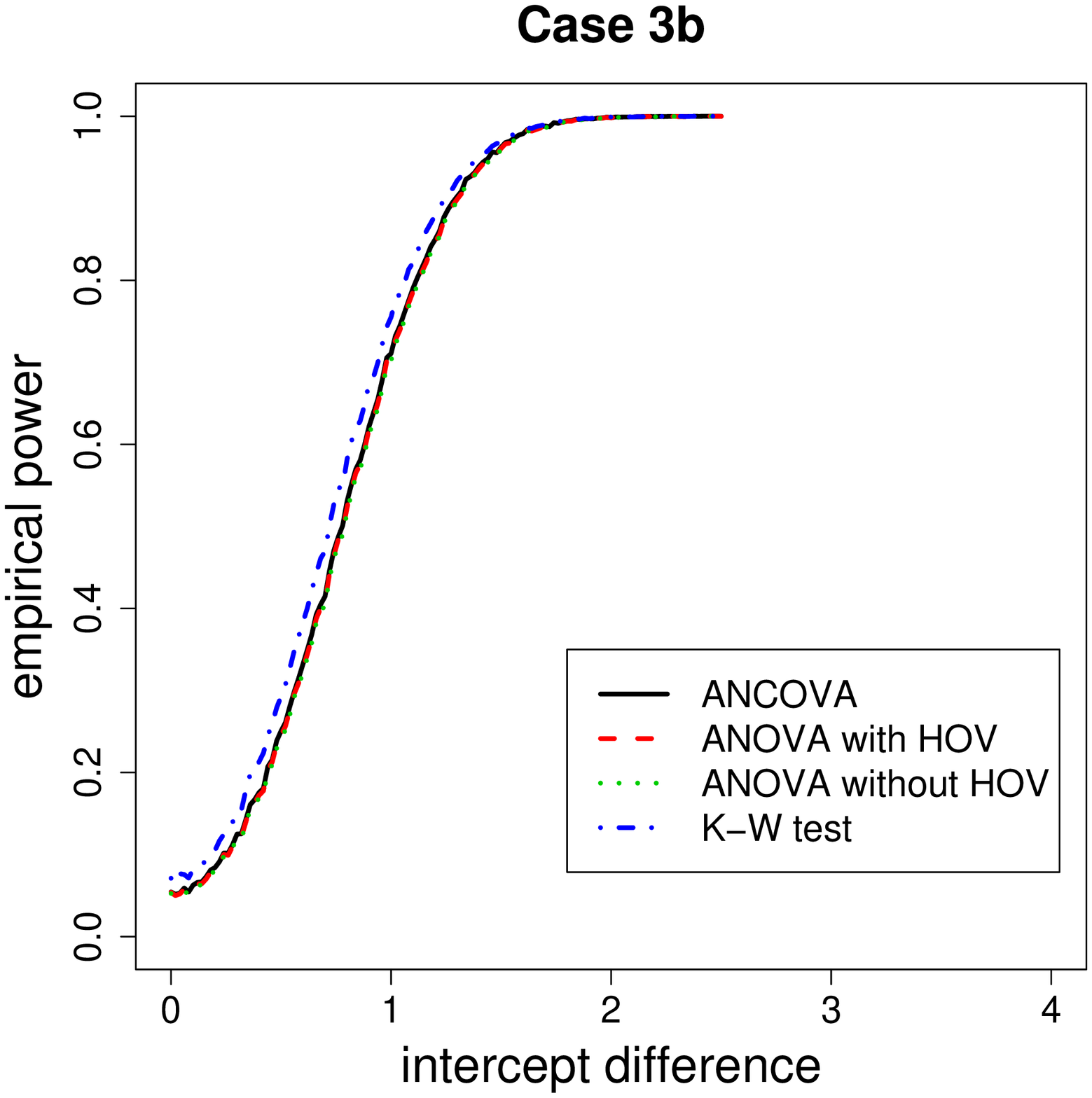}}}
\rotatebox{-90}{ \resizebox{3. in}{!}{ \includegraphics{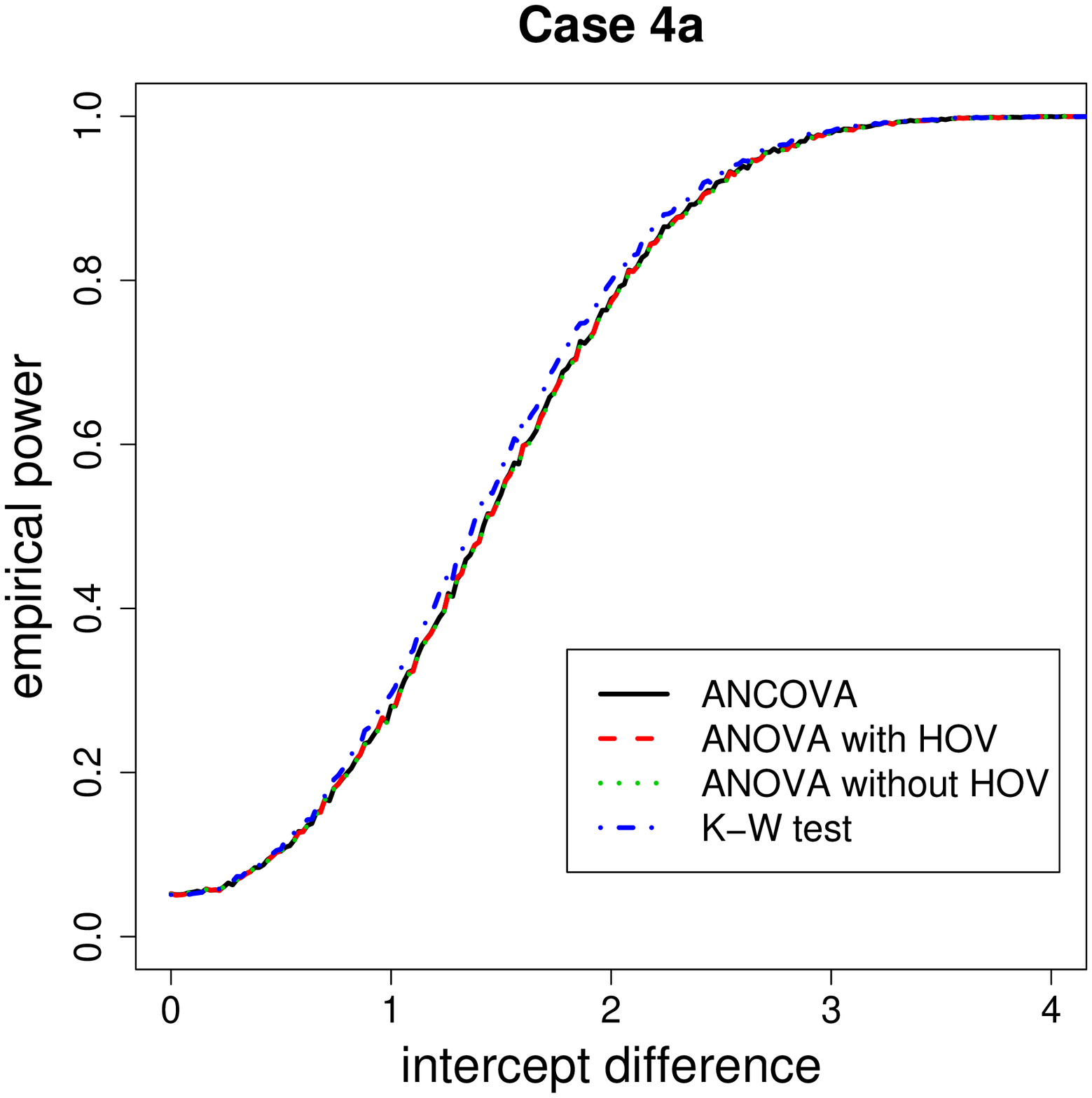}}}
\rotatebox{-90}{ \resizebox{3. in}{!}{ \includegraphics{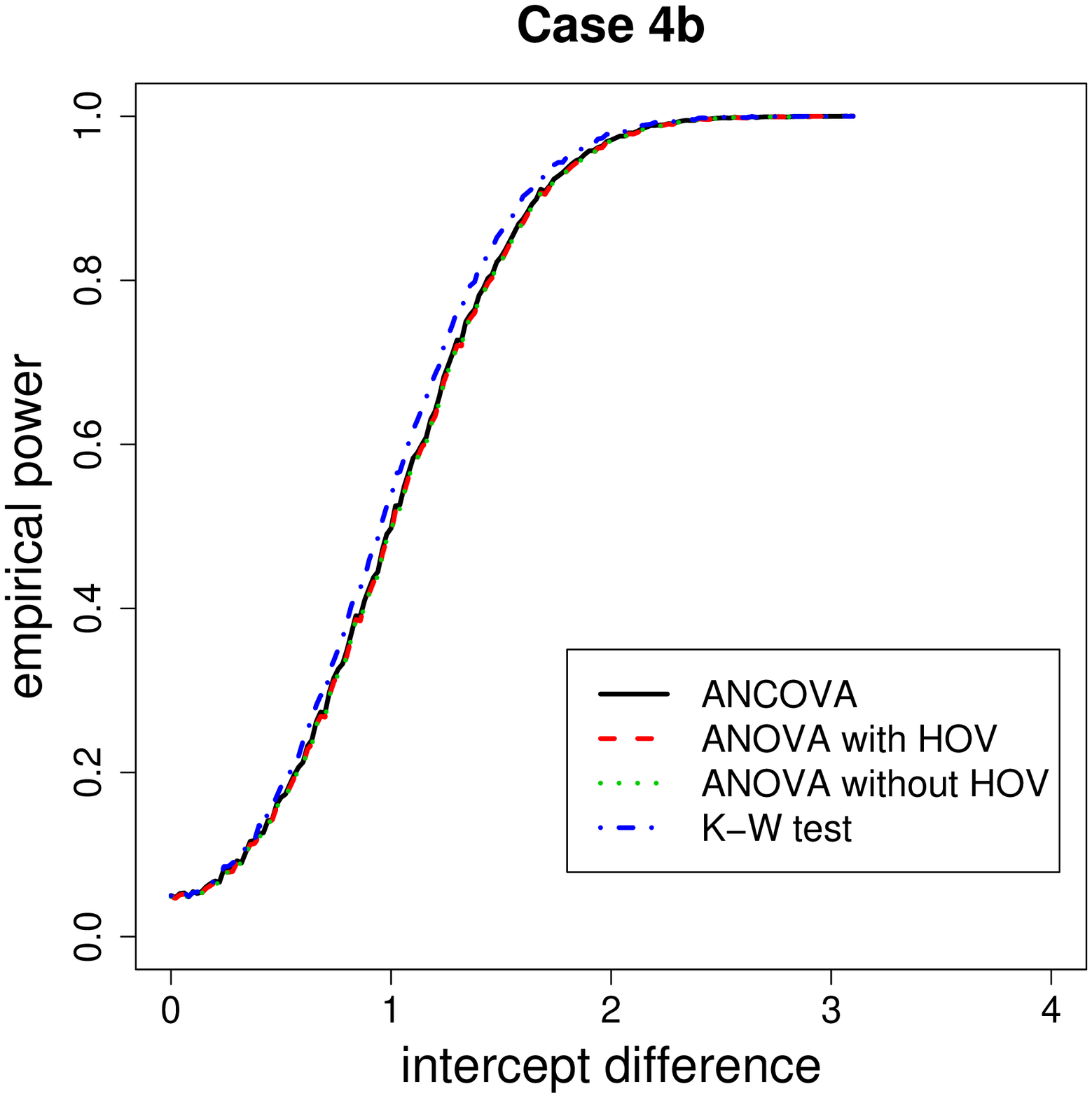}}}
\caption{
\label{fig:cases2-4}
Empirical power estimates versus intercept difference for cases 2a-4a, and 2b-4b.}
\end{figure}

\begin{figure}[htbp]
\centering
\rotatebox{-90}{ \resizebox{3. in}{!}{ \includegraphics{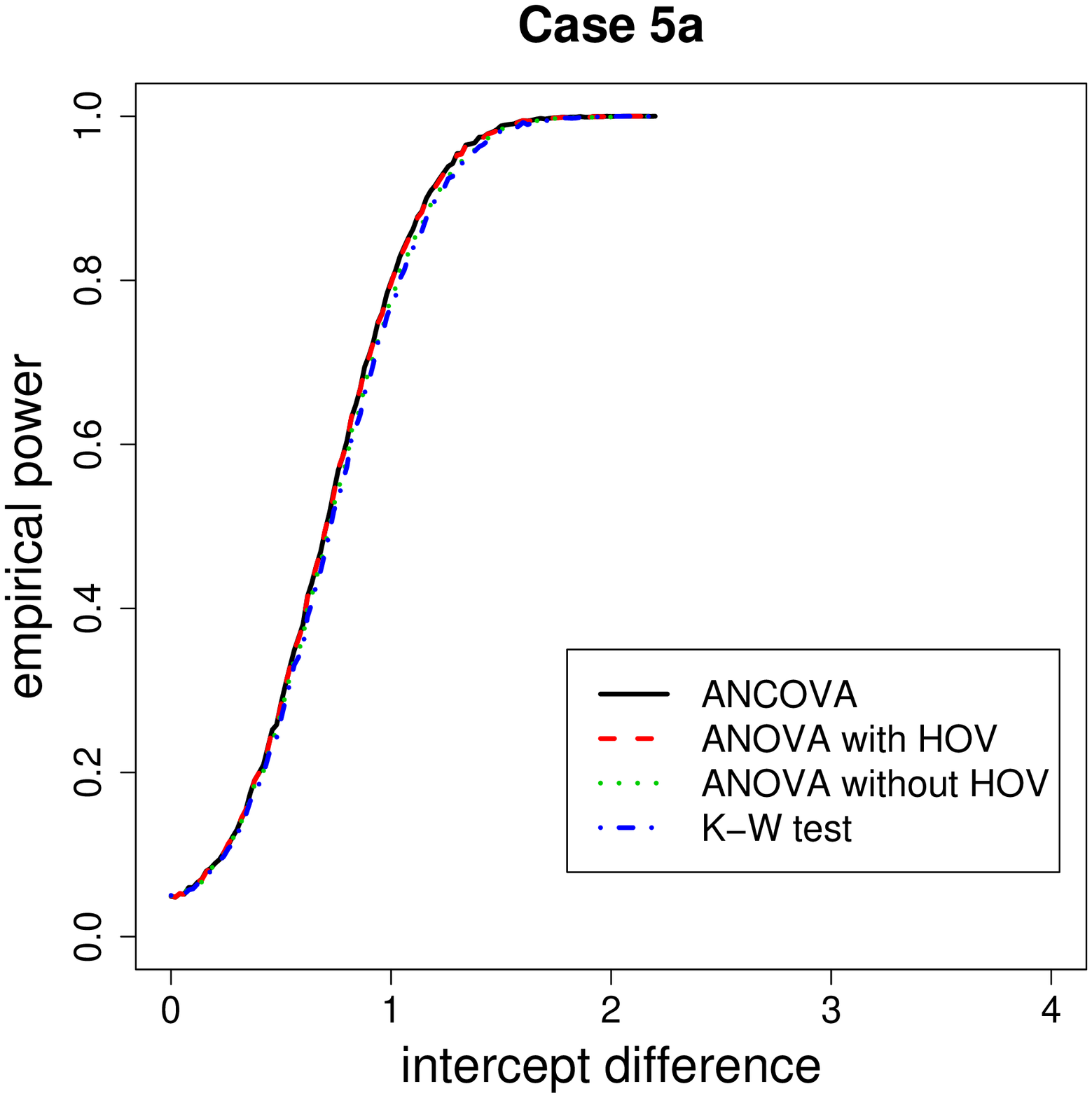}}}
\rotatebox{-90}{ \resizebox{3. in}{!}{ \includegraphics{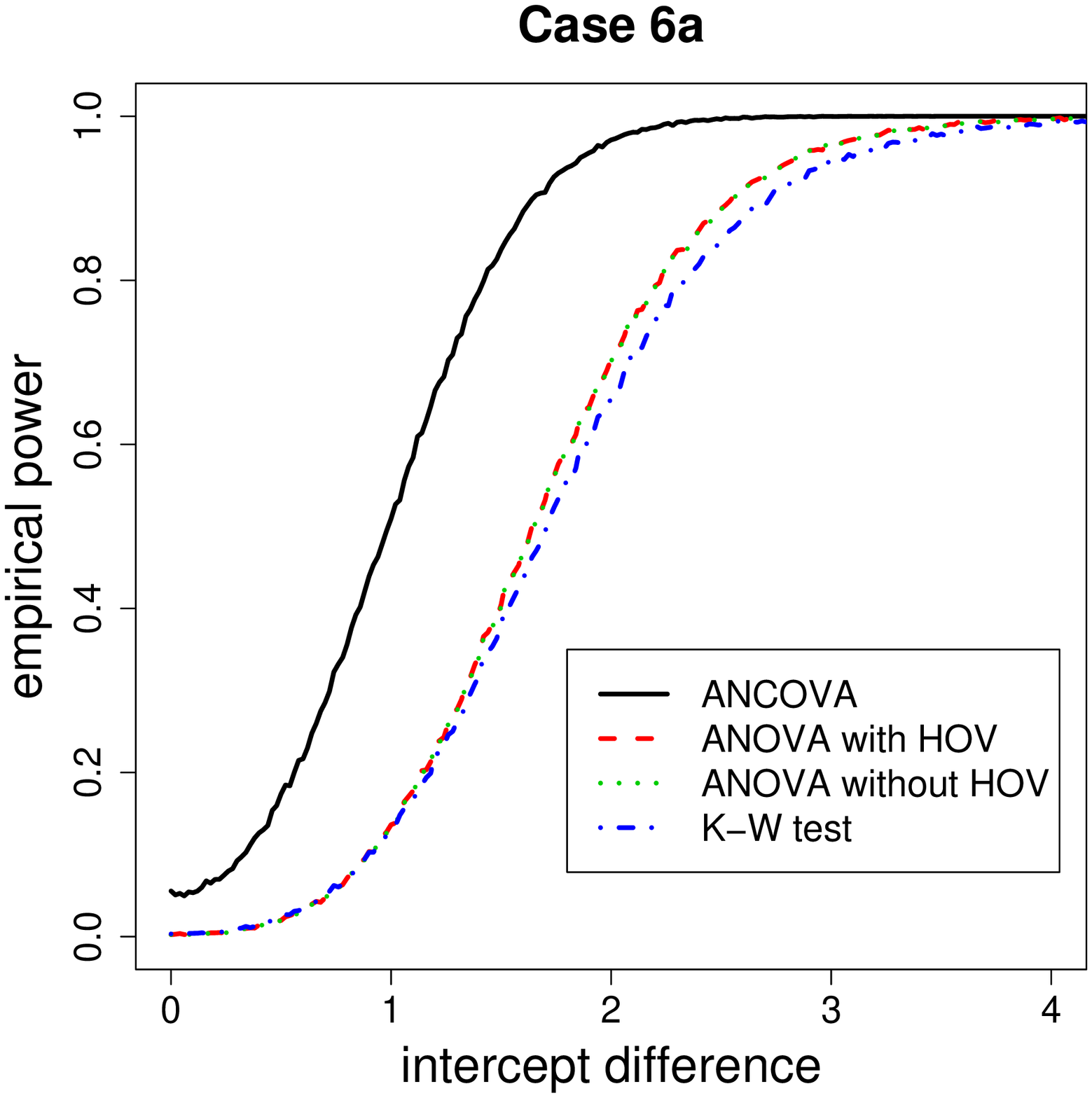}}}
\rotatebox{-90}{ \resizebox{3. in}{!}{ \includegraphics{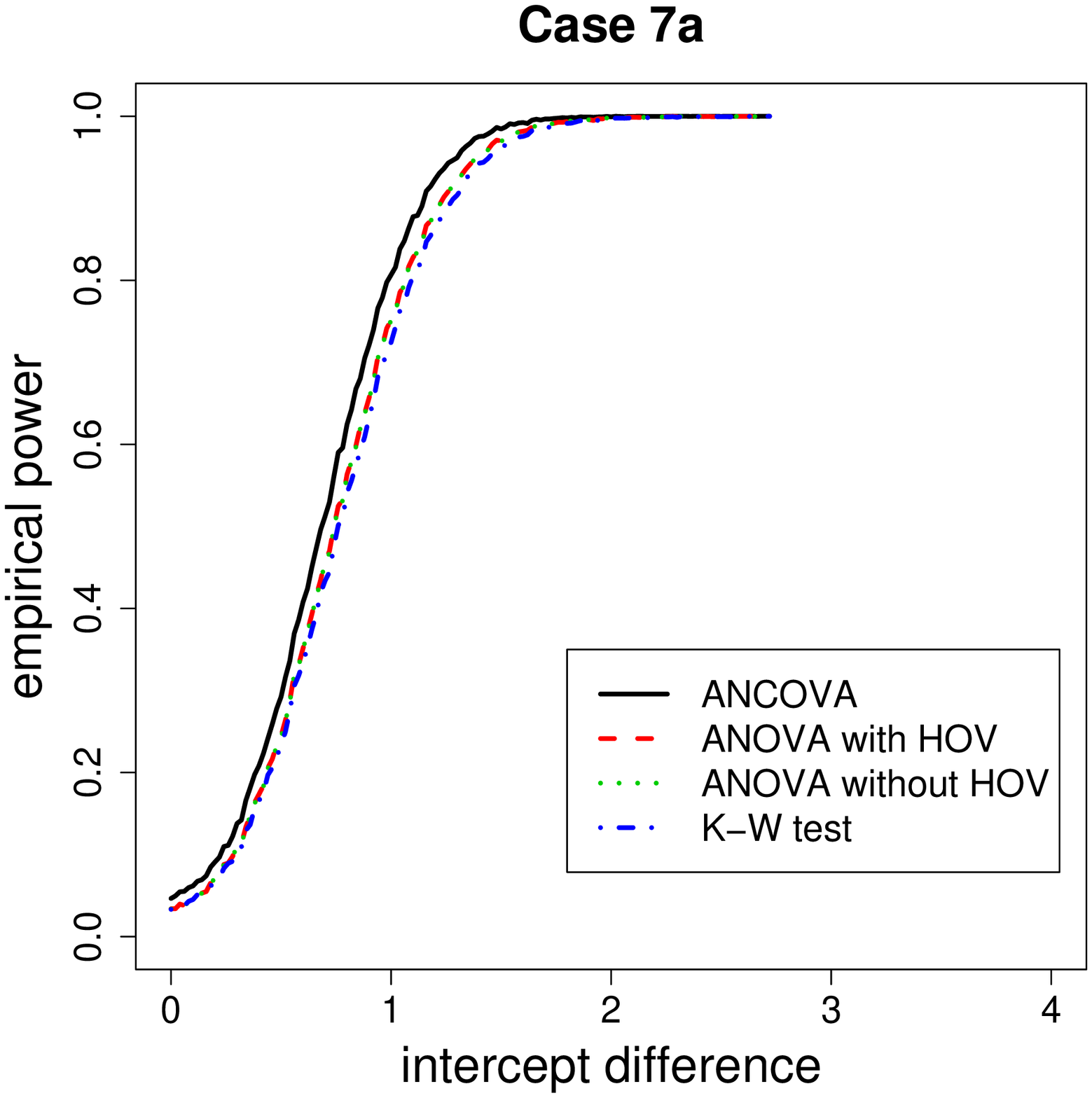}}}
\rotatebox{-90}{ \resizebox{3. in}{!}{ \includegraphics{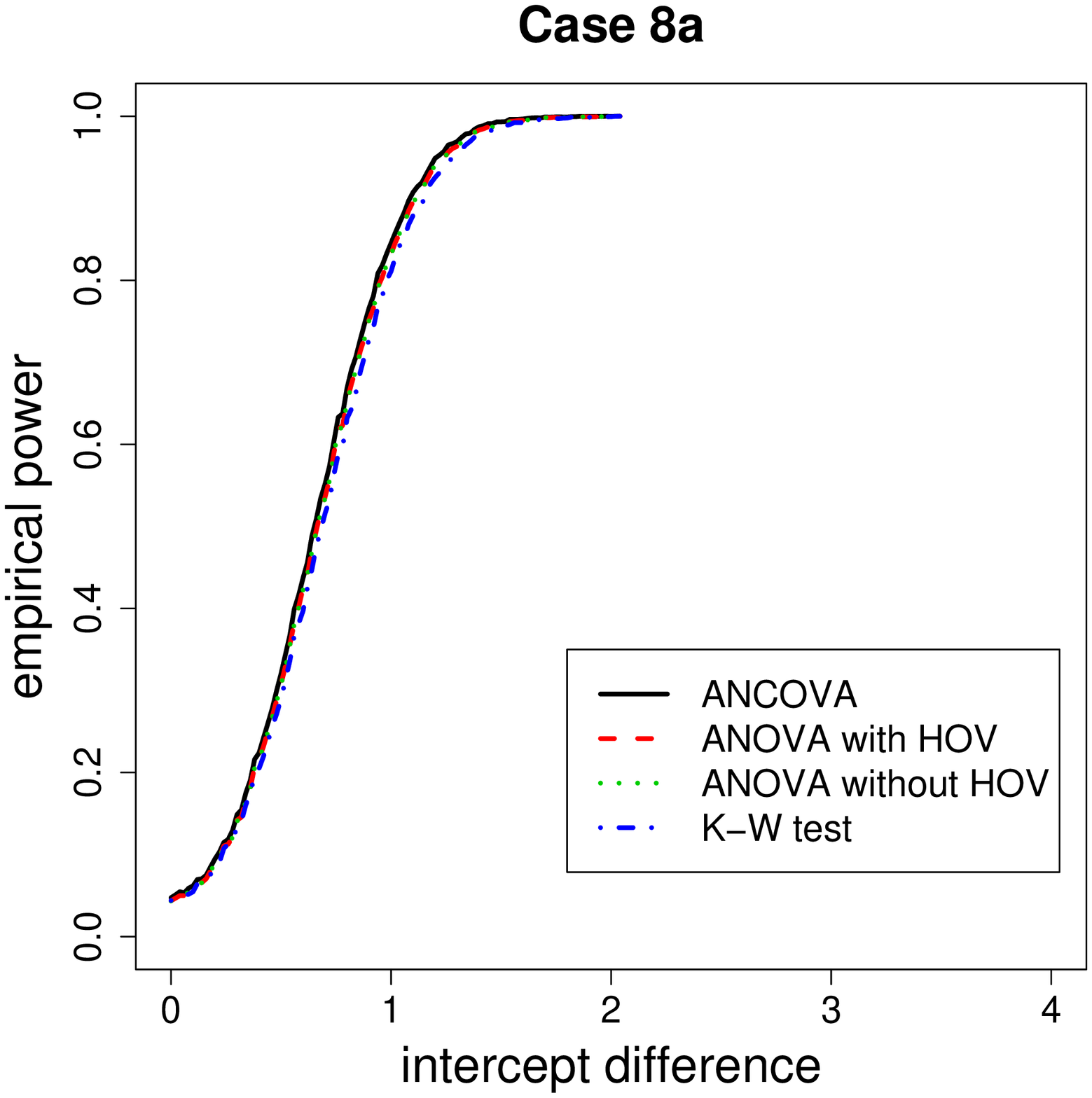}}}
\rotatebox{-90}{ \resizebox{3. in}{!}{ \includegraphics{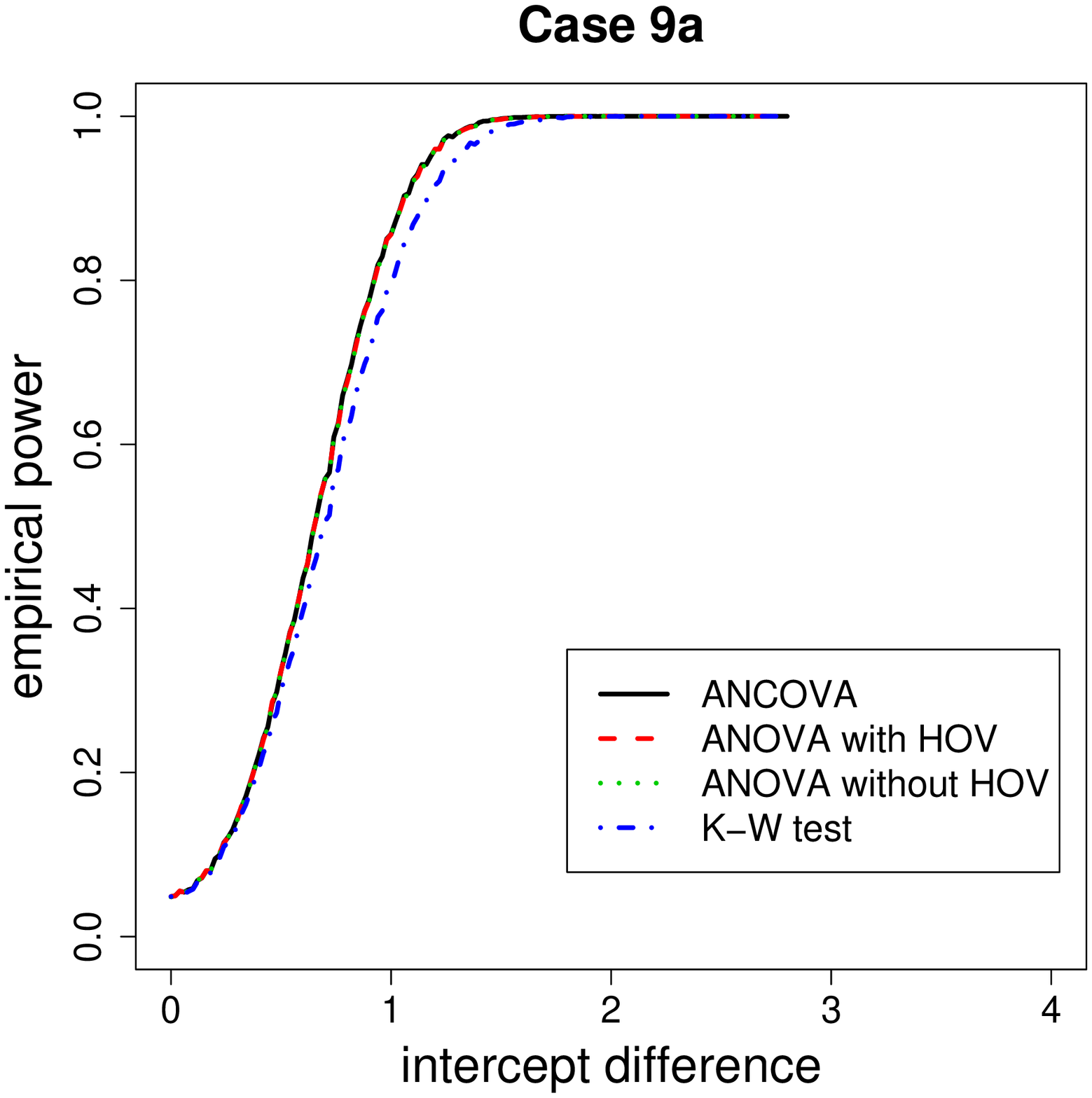}}}
\rotatebox{-90}{ \resizebox{3. in}{!}{ \includegraphics{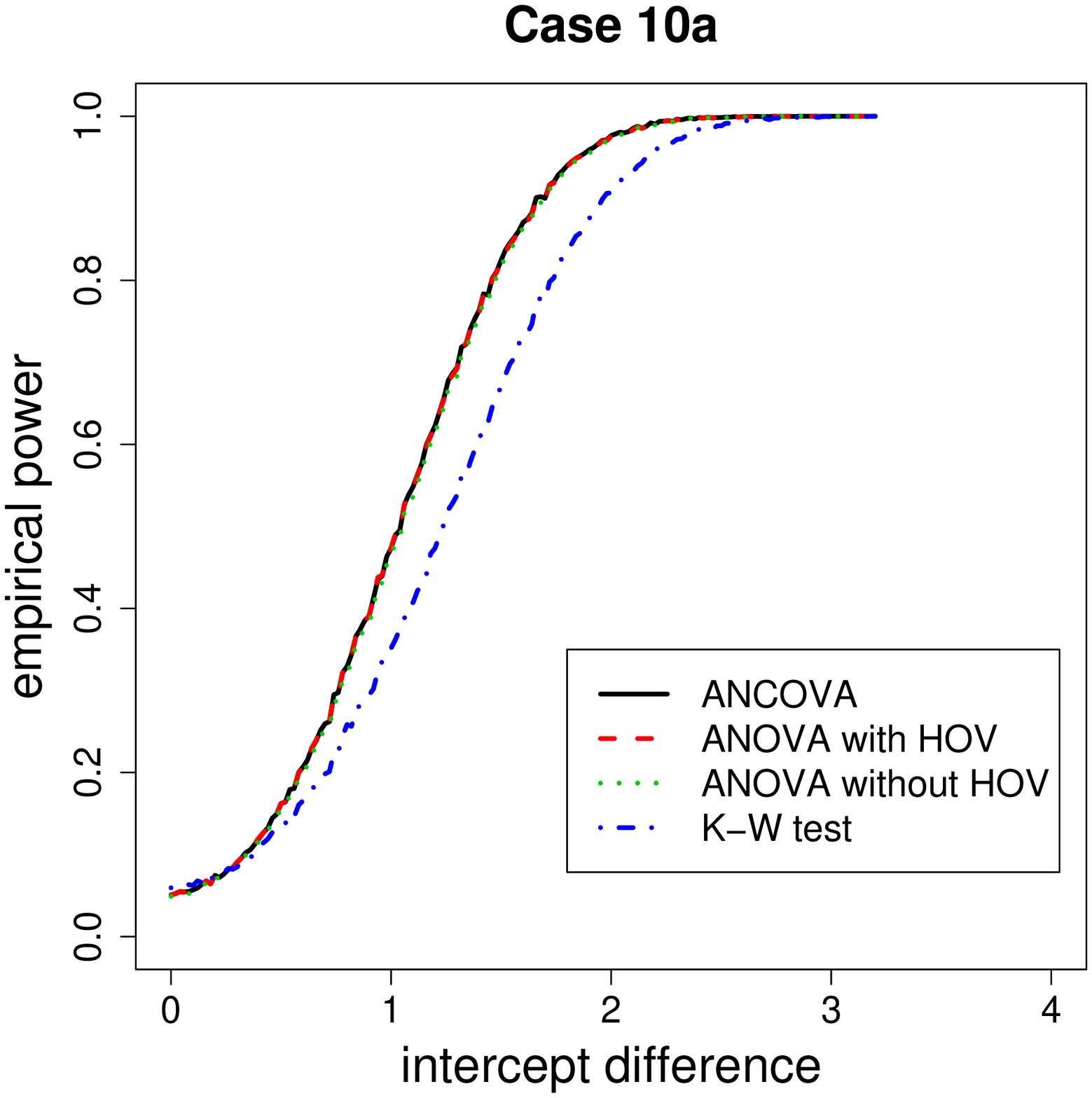}}}
\caption{
\label{fig:cases5-10}
Empirical power estimates versus intercept difference for cases 5a-10a.
}
\end{figure}

\begin{figure}[htbp]
\centering
\rotatebox{-90}{ \resizebox{3. in}{!}{ \includegraphics{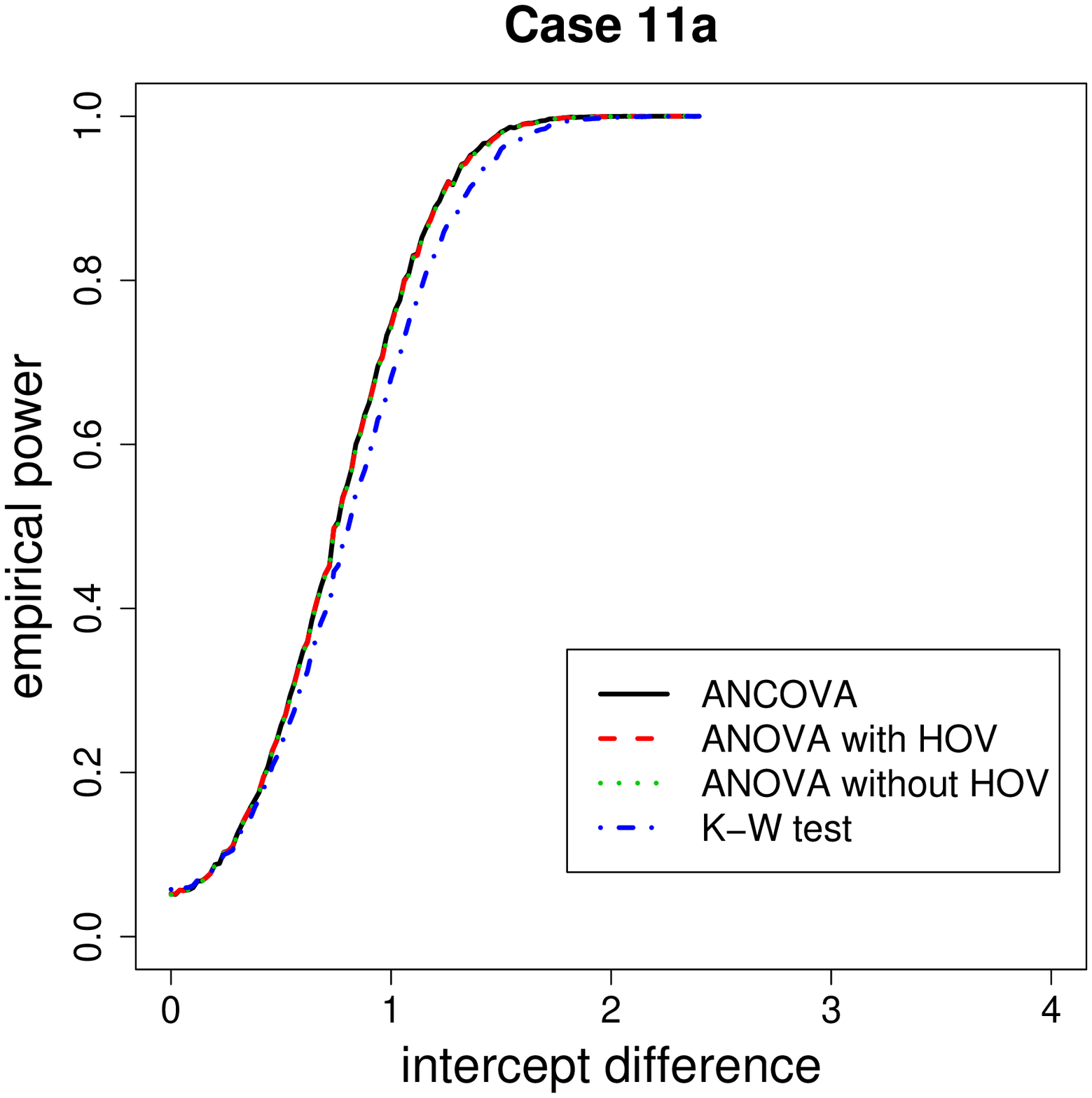}}}
\rotatebox{-90}{ \resizebox{3. in}{!}{ \includegraphics{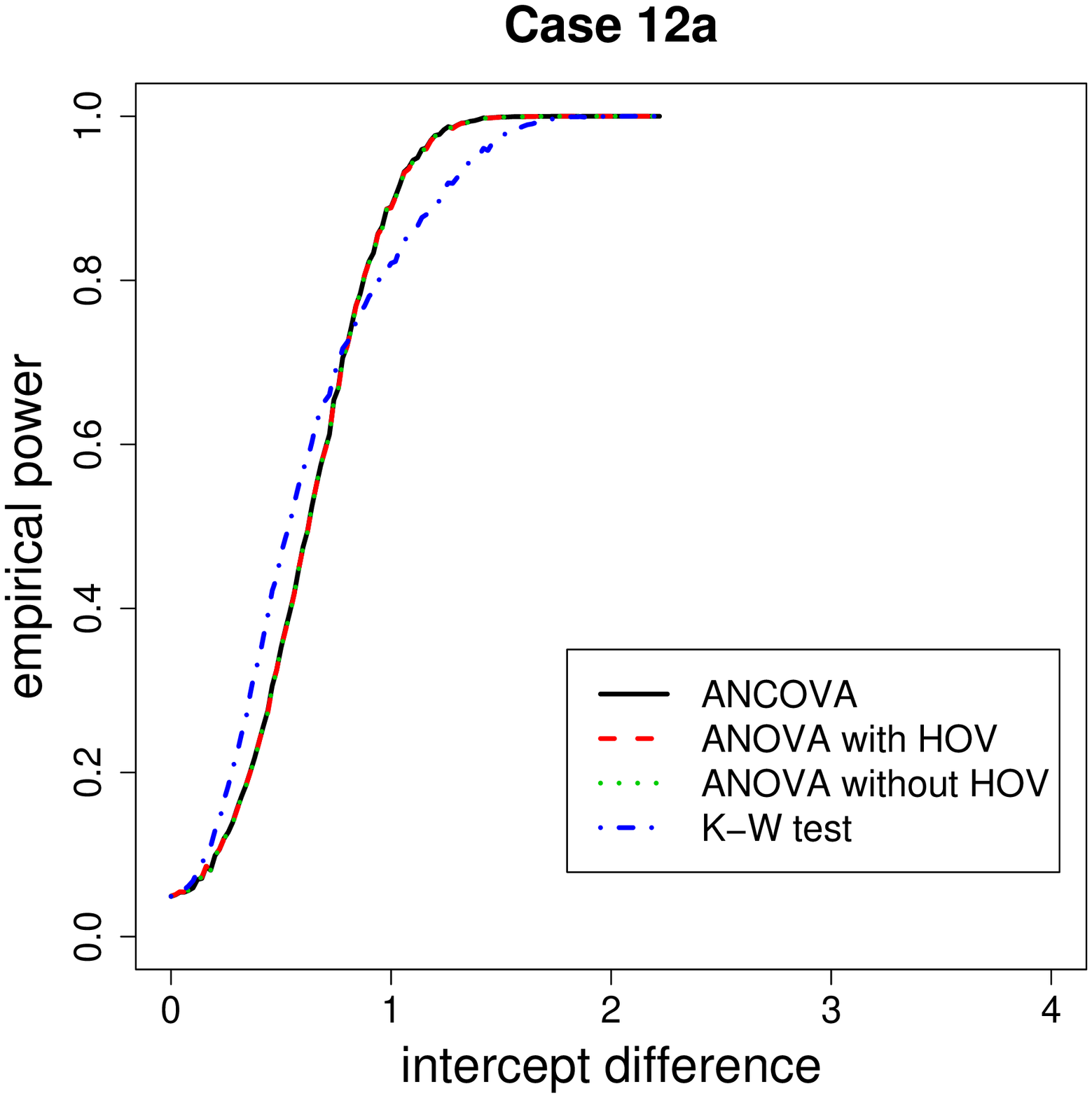}}}
\rotatebox{-90}{ \resizebox{3. in}{!}{ \includegraphics{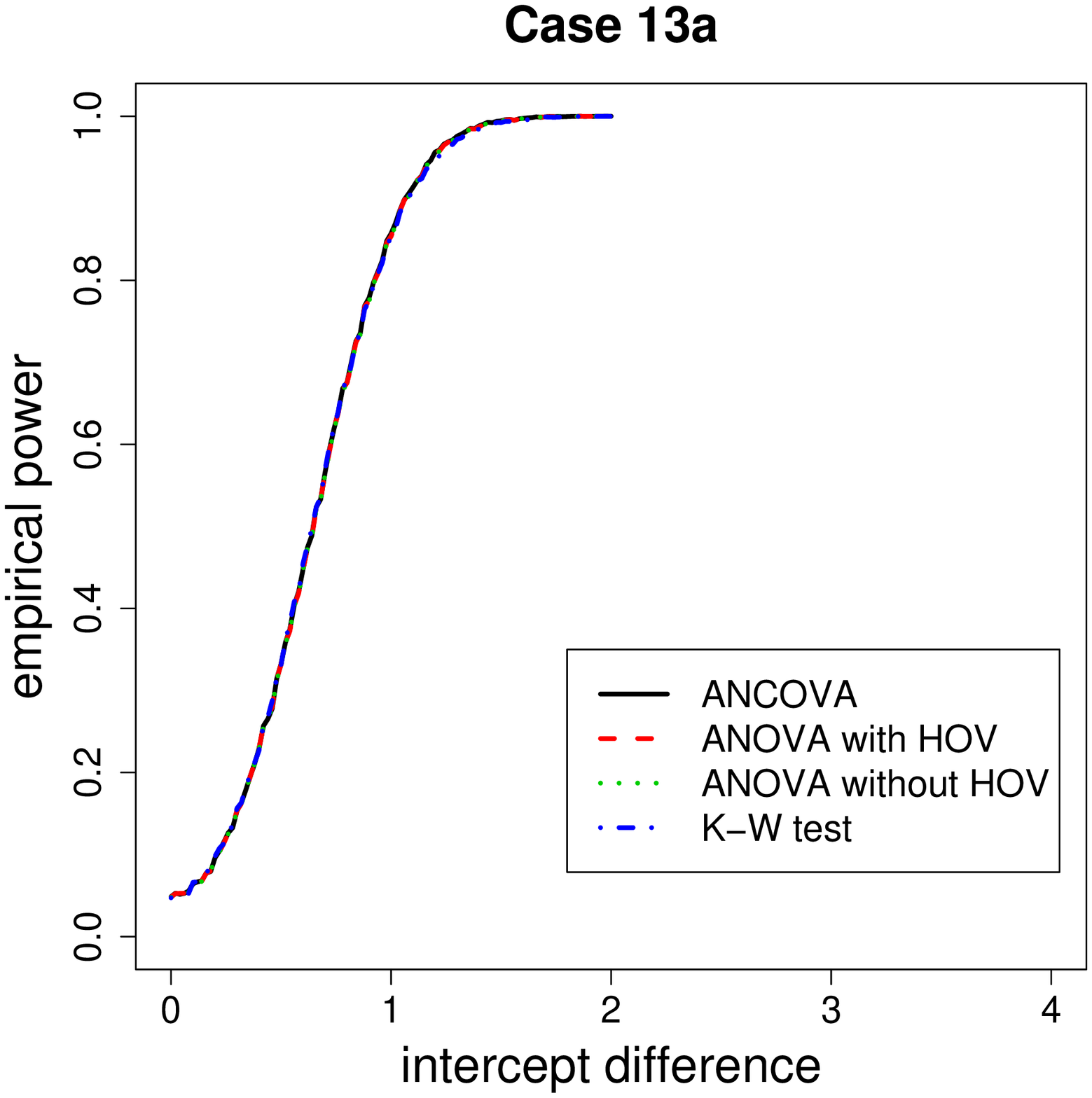}}}
\rotatebox{-90}{ \resizebox{3. in}{!}{ \includegraphics{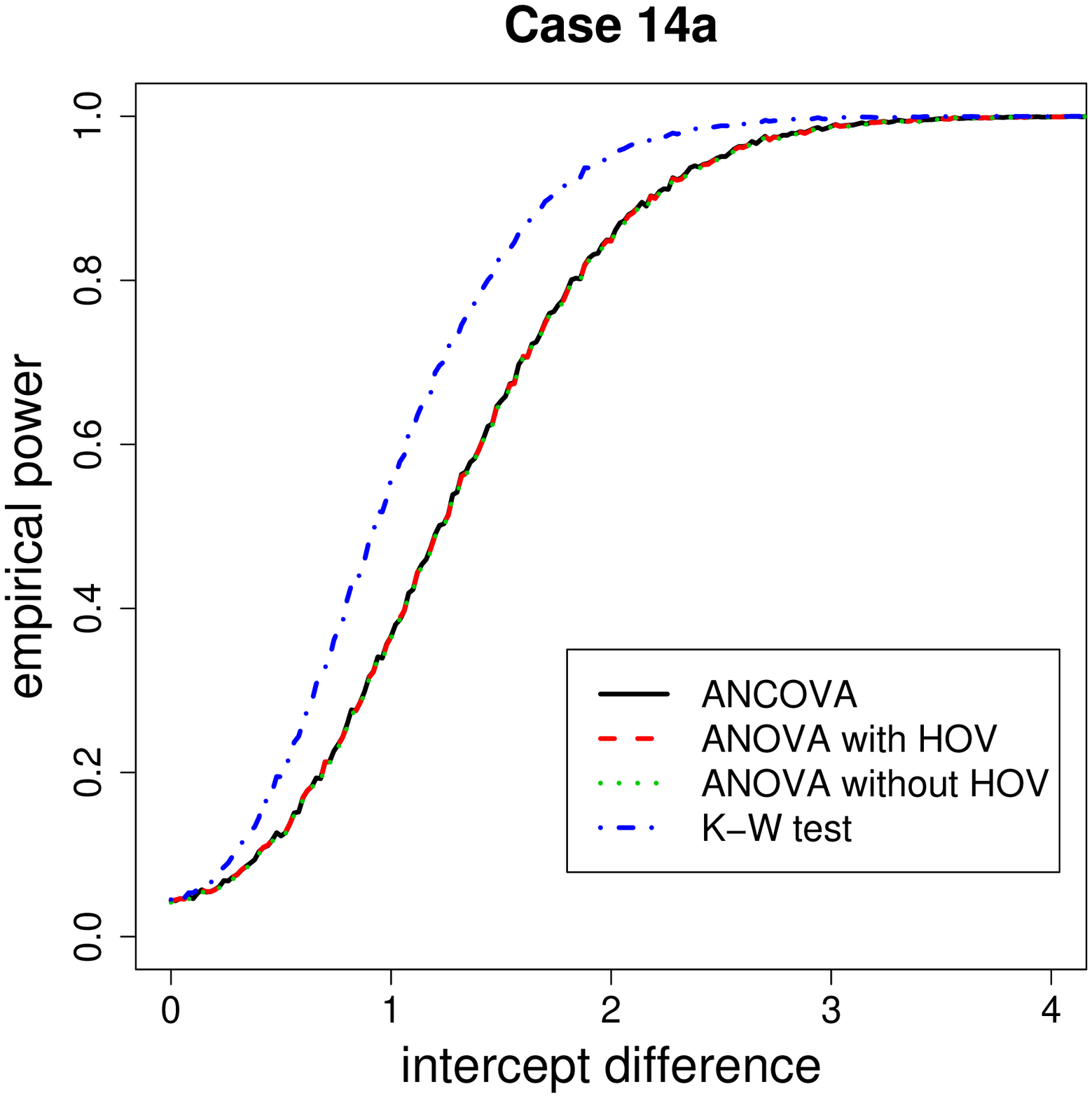}}}
\rotatebox{-90}{ \resizebox{3. in}{!}{ \includegraphics{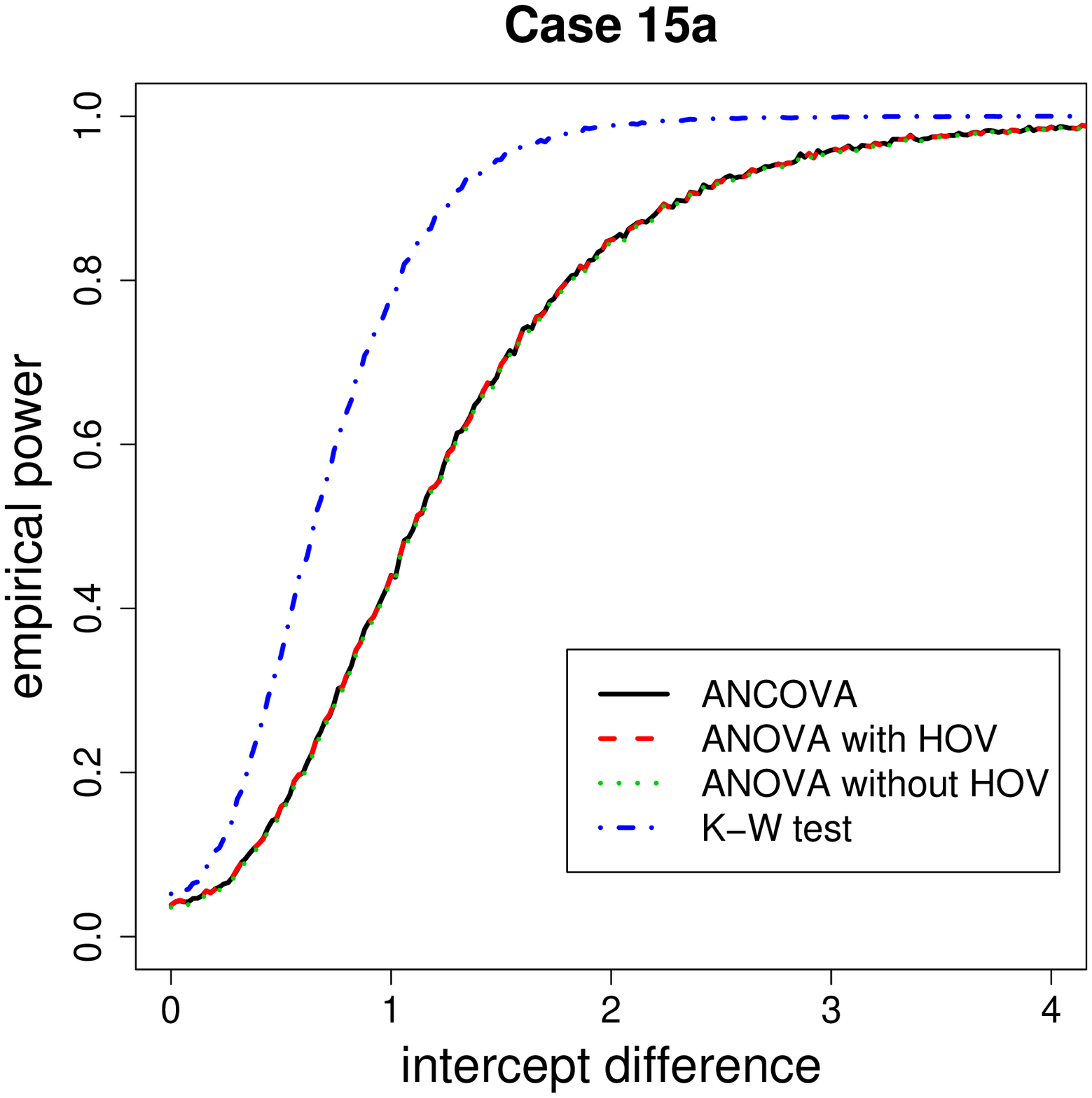}}}
\rotatebox{-90}{ \resizebox{3. in}{!}{ \includegraphics{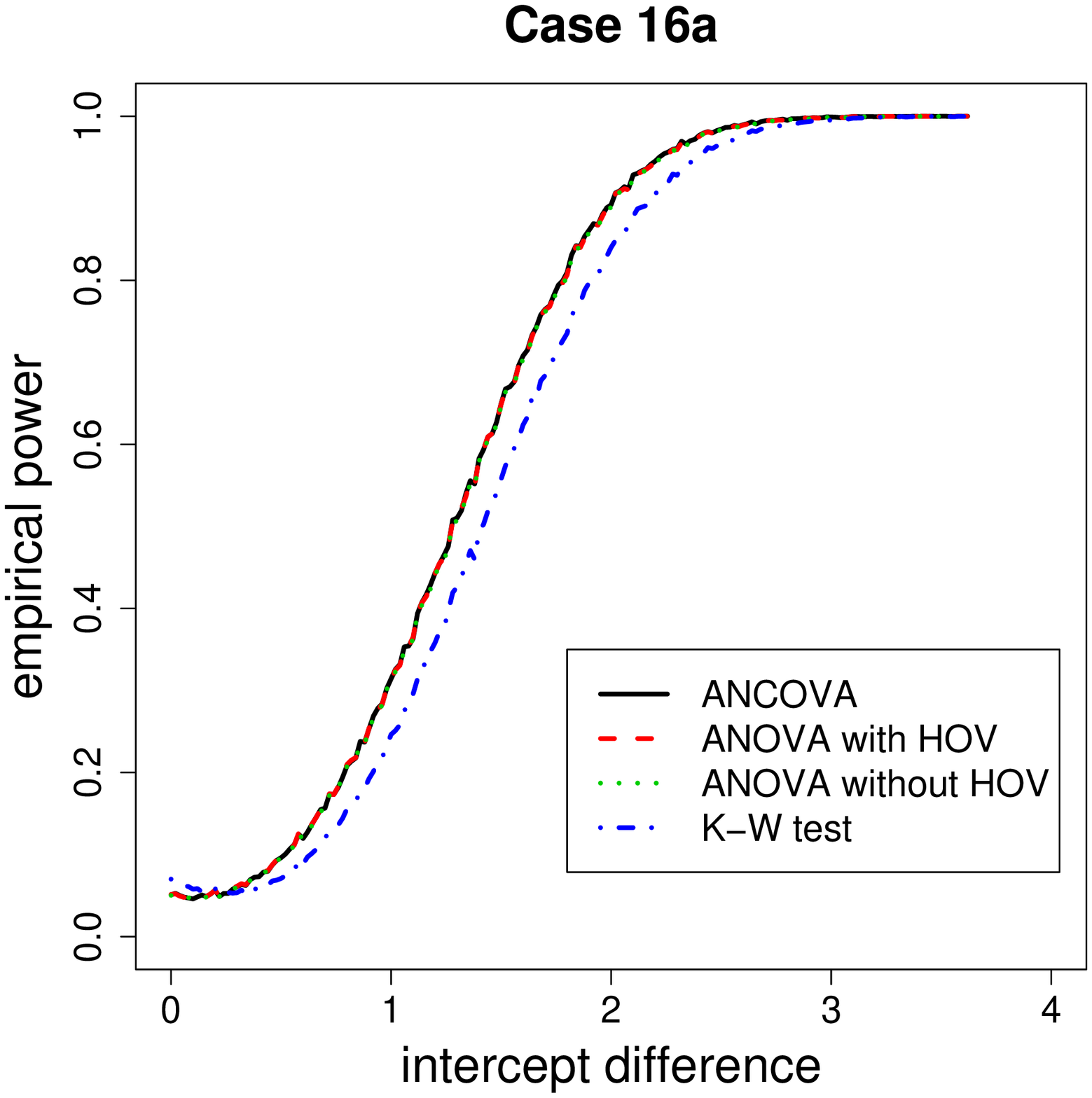}}}
\caption{
\label{fig:cases11-16}
Empirical power estimates versus intercept difference for cases 11a-16a.
}
\end{figure}

The empirical power curves are plotted in Figures \ref{fig:emp-power1-2a-b}, \ref{fig:cases2-4},
\ref{fig:cases5-10}, and \ref{fig:cases11-16}.
Empirical power corresponds to $\bh_i$, $i=1,2$.
The value on the horizontal axis is defined to be
intercept difference (i.e., $0.02\,q$) as in \eqref{eqn:generate-resp-trt2}.
Then the empirical power curves are plotted
against the simulated intercept difference values.
In these figures the empirical power curve for
a case labeled with ``a'' is steeper and approaches to 1.00 faster
than that of the case labeled with ``b'' for the same case number,
due to the fact that ``b''-labeled cases have two replicates with
the rest of the restrictions identical to the preceding ``a''-labeled cases.
Only cases labeled with ``a'' and ``b'' in case 1 are presented in Figure \ref{fig:emp-power1-2a-b}.
For other cases, plot for only ``a''-labeled case is presented.

The first intercept difference value at which the power reaches 1 are denoted as
$\kappa$ and are provided in Table \ref{tab:kappa-values} for all cases.
Observe also that power curves are steeper when error variances are smaller.
The empirical power curves are almost identical for all methods in case 13
which has a scaled Beta distribution for the error term.
That is, in this case the conditions balance out the power estimates for the methods.
In cases 1, 9-11, and 16 the power estimates for ANCOVA and ANOVA methods are similar
but all are larger than the K-W test power estimates.
In these cases, except in cases 11 and 16,
the error distributions are identical for both treatment levels,
and are all symmetric;
furthermore, uniform distribution approaching asymptotic normality
considerably fast satisfies all the assumptions of the parametric tests.
In cases 3, 4, 14, and 15 power estimates for ANCOVA and ANOVA methods are similar
but all are smaller than the K-W test power estimates.
In these cases, either HOV is violated as in cases 3 and 4,
or normality is violated as in cases 14 and 15 with the error distribution being asymmetric.
Since K-W test is non-parametric, it is robust to non-normality,
and since it tests distributional equality, it is more sensitive to HOV in normal cases.
In case 5, power estimates of ANCOVA and ANOVA with HOV are similar,
with both being larger than that of ANOVA without HOV
whose power estimate is larger than that of K-W test.
In this case,
the sample sizes for the treatments are different with everything else being same.
In cases 6-8, the power estimate of ANCOVA method is
significantly larger than those of the ANOVA methods whose empirical sizes
are larger than that of K-W test.
In these cases, the covariates are clustered with very different
treatment-specific means in cases 6 and 7, and similar means in case 8.
In cases 2 and 12, for smaller values of intercept difference
(i.e., between 0 to 0.5 in case 2 and 0 to 0.8 in case 12),
ANCOVA and ANOVA methods have similar power with all having a smaller power
than that of K-W test, while for larger values of the intercept difference
(i.e., between 0.5 to 4 in case 2 and 0.8 to 2 in case 12),
the order is reversed for the power estimates.
In case 2, error terms have different but constant variances,
and in case 12, error terms are non-normal but symmetric.

\section{Discussion and Conclusions}
\label{sec:disc-conc}
In this article,
we discuss various methods to remove the covariate influence on a response variable
when testing for differences between treatment levels.
The methods considered are the usual ANCOVA method and the
analysis of covariate-adjusted residuals using ANOVA with or without
homogeneity of variances (HOV) and Kruskal-Wallis (K-W) test.
The covariate-adjusted residuals are obtained from the fitted
overall regression line to the entire data set (ignoring the treatment levels).
For covariate-adjusted residuals to be appropriate for removing the covariate influence,
the treatment-specific lines and the overall regression line should be parallel.
On the other hand, ANCOVA can be used to test the equality of treatment means at specific values of the covariate.
Furthermore, the use of ANCOVA is extended to the nonparallel treatment-specific
lines also (\cite{kowalski:1994}).

The Monte Carlo simulations indicate that
when the covariates have similar means and have similar
distributions (with or without HOV),
ANCOVA, ANOVA with or without HOV methods have similar empirical sizes;
and K-W test is sensitive to distributional differences,
since the null hypotheses for the first three tests are about same
while it is more general for K-W test.
When the treatment-specific lines are parallel,
treatment-specific covariate ranges and covariate distributions are similar.
ANCOVA and ANOVA with or without HOV on covariate-adjusted residuals give similar results
if error variances have symmetric distributions with or without HOV and
sample sizes are similar for treatments;
give similar results if error variances are homogeneous
and sample sizes are different but large for treatments.
In these situations,
parametric tests are more powerful than K-W test.
The methods give similar results
but are liberal if error variances are heterogeneous
with different functional forms for treatments.
In these cases, usually K-W test has better performance.

When the treatment-specific lines are parallel,
but treatment-specific covariate ranges are different;
i.e., there exist clustering of the covariate relative to the treatment factors,
ANCOVA and ANOVA on covariate-adjusted residuals yield similar results
if treatment-specific covariate means are similar,
very different results if treatment-specific covariate means are different
since overall regression line will not be parallel to the treatment-specific lines.
In such a case,
methods on covariate-adjusted residuals tend to be extremely conservative
whereas the size of ANCOVA $F$ test is about the desired nominal level.
Moreover, ANCOVA is much more powerful than ANOVA on covariate-adjusted residuals in these cases.
The power of ANOVA on covariate-adjusted residuals gets closer to that of ANCOVA,
as the difference between the treatment-specific covariate means gets smaller.
However, in the case of clustering of covariates relative to the treatments,
one should also exercise extra caution due to the extrapolation problem.
Moreover in practice, such clustering is suggestive of an ignored grouping factor as in blocking.
The discussed methods are meaningful only within the
overlap of the clusters or in the close vicinity of them.
However, when there are clusters for the groups in terms of the covariate,
it is very likely that covariate and the group factors are dependent,
which violates an assumption for ANCOVA.
When this dependence is strong then ANCOVA method will not be appropriate.
On the other hand, the residual analysis is extremely conservative
which might be viewed as an advantage
in order not to reach spurious and confounded conclusions in such a case.

The ANCOVA models can be used to estimate the treatment-specific
response means at specific values of the covariate.
But the ANOVA model on covariate-adjusted residuals should be used together
with the fitted overall regression line in such an estimation,
as long as condition \eqref{eqn:null-ANCOVA-equal-intercept} holds.

Different treatment-specific covariate distributions
within the same interval or different intervals
might also cause treatment-specific covariate means to be different.
In such a case,
ANCOVA should be preferred against the methods on covariate-adjusted residuals.

In conclusion, we recommend the following strategy for the use of the above methods:
(i) First, one should check the significance of the effect of the covariates for each treatment,
i.e., test $H_o^i:$ ``all treatment-specific slopes are equal to zero".
If $H_o^i$ is not rejected, then the usual (one-way) ANOVA or K-W test can be used.
(ii) If $H_o^i$ is rejected, the covariate effect is significant for at least one treatment factor.
Hence one should test $H_o^{ii}:$ ``equality of all treatment-specific slopes".
If $H_o^{ii}$ is rejected, then the covariate should be included in the analysis
as an important variable and the usual regression tools can be employed.
(iii) If $H_o^{ii}$ is not rejected, check the covariate ranges.
If they are similar or have a considerable intersection for treatment factors,
then ANCOVA and methods on residuals are appropriate.
Then one should check the underlying assumptions for the methods
and then pick the best method among them.
(iv) If covariate ranges are very different,
then it is very likely that treatment and covariate are not independent,
hence ANCOVA is not appropriate.
On the other hand, the methods on residuals can be used but they are extremely conservative.
In this case, one may apply some other method,
e.g., MANOVA on (response,covariate) data for treatment differences.



\begin{thebibliography}{}

\bibitem[Akritas et~al., 2000]{akritas:2000}
Akritas, M., Arnold, S., and Du, Y. (2000).
\newblock Nonparametric models and methods for nonlinear analysis of
  covariance.
\newblock {\em Biometrika}, 87(3):507--526.

\bibitem[Albrecht et~al., 1993]{albrecht:1993}
Albrecht, G.~H., Gelvin, B.~R., and Hartman, S.~E. (1993).
\newblock Ratios as a size adjustment in morphometrics.
\newblock {\em American Journal of Physical Anthropology}, 91(4):441--468.

\bibitem[Atchley et~al., 1976]{atchley:1976}
Atchley, W.~R., Gaskins, C.~T., and Anderson, D. (1976).
\newblock Statistical properties of ratios {I. E}mpirical results.
\newblock {\em Systematic Zoology}, 25(2):137--148.

\bibitem[Beaupre and Duvall, 1998]{beaupre:1998}
Beaupre, S.~J. and Duvall, D. (1998).
\newblock Variation in oxygen consumption of the western diamondback
  rattlesnake ({C}rotalus atrox): implications for sexual size dimorphism.
\newblock {\em Journal Journal of Comparative Physiology {B}: Biochemical,
  Systemic, and Environmental Physiology}, 168(7):497--506.

\bibitem[Ceyhan, 2000]{ceyhan:masters-thesis}
Ceyhan, E. (2000).
\newblock A comparison of analysis of covariance and {ANOVA} methods using
  covariate-adjusted residuals.
\newblock Master's thesis, Oklahoma State University, Stillwater, OK, 74078.

\bibitem[Garcia-Berthou, 2001]{garcia-berthou:2001}
Garcia-Berthou, E. (2001).
\newblock On the misuse of residuals in ecology: Testing regression residuals
  vs. the analysis of covariance.
\newblock {\em The Journal of Animal Ecology}, 70(4):708--711.

\bibitem[Huitema, 1980]{huitema:1980}
Huitema, B.~E. (1980).
\newblock {\em The Analysis of Covariance and Alternatives}.
\newblock John Wiley \& Sons Inc., New York.

\bibitem[Jakob et~al., 1996]{jakob:1996}
Jakob, E.~M., Marshall, S.~D., and Uetz, G.~W. (1996).
\newblock Estimating fitness: A comparison of body condition indices.
\newblock {\em Oikos}, 77(1):61--67.

\bibitem[Kowalski et~al., 1994]{kowalski:1994}
Kowalski, C.~J., Schneiderman, E.~D., and Willis, S.~M. (1994).
\newblock {ANCOVA} for nonparallel slopes: the {Johnson-Neyman} technique.
\newblock {\em International Journal of Bio-Medical Computing}, 37(3):273--286.

\bibitem[Kuehl, 2000]{kuehl:2000}
Kuehl, R.~O. (2000).
\newblock {\em Design of Experiments: Statistical Principles of Research Design
  and Analysis. (2nd ed.)}.
\newblock Pacific Grove, CA: Brooks/Cole.

\bibitem[Kutner et~al., 2004]{kutner:2004}
Kutner, M.~H., Nachtsheim, C.~J., and Neter, J. (2004).
\newblock {\em Applied Linear Regression Models. (4th ed.)}.
\newblock McGraw-Hill/Irwin, Chicago, IL.

\bibitem[Maxwell et~al., 1984]{maxwell:1984}
Maxwell, S.~E., Delaney, H.~D., and Dill, C.~A. (1984).
\newblock Another look at {ANCOVA} versus blocking.
\newblock {\em Psychological Bulletin}, 95(1):136–147.

\bibitem[Maxwell et~al., 1985]{maxwell:1985}
Maxwell, S.~E., Delaney, H.~D., and Manheimer, J.~M. (1985).
\newblock {ANOVA} of residuals and {ANCOVA}: Correcting an illusion by using
  model comparisons and graphs.
\newblock {\em Journal of Educational Statistics}, 10(3):197--209.

\bibitem[Miller and Chapman, 2001]{miller-ANCOVA:2001}
Miller, G.~A. and Chapman, J.~P. (2001).
\newblock Misunderstanding analysis of covariance.
\newblock {\em Journal of Abnormal Psychology}, 110(1):40--48.

\bibitem[Milliken and Johnson, 2002]{milliken:2002}
Milliken, G. and Johnson, D.~E. (2002).
\newblock {\em Analysis of Messy Data, Volume {III}: Analysis of Covariance}.
\newblock Chapman and Hall/CRC, New York.

\bibitem[Ott, 1993]{ott:1993}
Ott, R.~L. (1993).
\newblock {\em An Introduction to Statistical Methods and Data Analysis. (4th
  ed.)}.
\newblock Duxbury Press, Belmont, CA.

\bibitem[Packard and Boardman, 1988]{packard:1988}
Packard, G.~C. and Boardman, T.~J. (1988).
\newblock The misuse of ratios, indices, and percentages in ecophysiological
  research.
\newblock {\em Physiological Zoology}, 61:1--9.

\bibitem[Raubenheimer and Simpson, 1992]{raubenheimer:1992}
Raubenheimer, D. and Simpson, S.~J. (1992).
\newblock Analysis of covariance: an alternative to nutritional indices.
\newblock {\em Journal Entomologia Experimentalis et Applicata},
  62(3):221--231.

\bibitem[Rheinheimer and Penfield, 2001]{rheinheimer:2001}
Rheinheimer, D.~C. and Penfield, D.~A. (2001).
\newblock The effects of type {I} error rate and power of the {ANCOVA $F$} test
  and selected alternatives under nonnormality and variance heterogeneity.
\newblock {\em Journal of Experimental Education}, 69(4):373--391.

\bibitem[Siegel and Castellan~Jr., 1988]{siegel:1988}
Siegel, S. and Castellan~Jr., N.~J. (1988).
\newblock {\em Nonparametric Statistics for the Behavioral Sciences (second
  edition)}.
\newblock McGraw-Hill, New York.

\bibitem[Small, 1996]{small:1996}
Small, C.~G. (1996).
\newblock {\em The Statistical Theory of Shape}.
\newblock Springer-Verlag, New York.

\bibitem[Tsangari and Akritas, 2004a]{tsangari:2004JMVA}
Tsangari, H. and Akritas, M.~G. (2004a).
\newblock Nonparametric {ANCOVA} with two and three covariates.
\newblock {\em Journal of Multivariate Analysis}, 88(2):298--319.

\bibitem[Tsangari and Akritas, 2004b]{tsangari:2004JNPS}
Tsangari, H. and Akritas, M.~G. (2004b).
\newblock Nonparametric models and methods for ancova with dependent data.
\newblock {\em Journal of Nonparametric Statistics}, 16(3-4):403--420.

\end{thebibliography}

\section{Tables}
\label{sec:tables}

\begin{table}[htp]
\centering
\begin{tabular}{|c|c|c|c|c|c|c|}
\hline
&
\multicolumn{2}{|c|}{error term} &
\multicolumn{2}{|c|}{sample sizes} &
\multicolumn{2}{|c|}{ranges of covariate for}  \\
\hline
case & $e_{1jk} \stackrel{ind}{\sim}$ & $e_{2jk} \stackrel{ind}{\sim}$ & $n_1 $ & $n_2 $ & treatment 1& treatment 2  \\
\hline
1 & $N(0,1)$ & $N(0,1)$& 20& 20& (0,10)& (0,10) \\
\hline
2 & $N(0,1)$ & $N\left({0,6} \right)$& 20& 20& (0,10)& (0,10) \\
\hline
3 & $N(0,1)$ & $N(0,\sqrt{x})$& 20& 20& (0,10)& (0,10) \\
\hline
4 & $N(0,\sqrt{x})$ & $N(0,\sqrt{x})$& 20& 20& (0,10)& (0,10) \\
\hline
5 & $N(0,1)$ & $N(0,1)$& 28& 12& (0,10)& (0,10) \\
\hline
6 & $N(0,1)$ & $N(0,1)$& 20& 20& (0,6)& (4,10) \\
\hline
7 & $N(0,1)$ & $N(0,1)$& 20& 20& (0,3)$\cup $(7,10)& (4,10) \\
\hline
8 & $N(0,1)$ & $N(0,1)$& 20& 20& (0,4)$\cup $(6,10)& (3,7) \\
\hline
9 & $\U\left({-\sqrt 3,\sqrt 3} \right)$ & $\U\left({-\sqrt 3,\sqrt 3} \right)$& 20& 20& (0,10)& (0,10) \\
\hline
10 & $\U\left({-\sqrt 3,\sqrt 3} \right)$ & $\U\left({-2\sqrt 3,2\sqrt 3} \right)$& 20& 20& (0,10)& (0,10) \\
\hline
11 & $\U\left({-\sqrt 3,\sqrt 3} \right)$ & $\U\left({-\sqrt{x},\sqrt{x}} \right)$& 20& 20& (0,10)& (0,10) \\
\hline
12 & $DW\left({0,1,3} \right)$ & $DW\left({0,1,3} \right)$& 20& 20& (0,10)& (0,10) \\
\hline
13 & $\sqrt{48} \left({\be \left({6,2} \right)-3 \mathord{\left/{\vphantom{3 4}} \right. \kern-\nulldelimiterspace} 4} \right)$&
$\sqrt{48} \left({\be \left({6,2} \right)-3 \mathord{\left/{\vphantom{3 4}} \right. \kern-\nulldelimiterspace} 4} \right)$&
20& 20& (0,10)& (0,10) \\
\hline
14 & $\chi_2^2 -2$ & $\chi_2^2 -2$& 20& 20& (0,10)& (0,10) \\
\hline
15 & $LN(0,1)-e^{1 \mathord{\left/{\vphantom{1 2}} \right. \kern-\nulldelimiterspace} 2}$&
$LN(0,1)-e^{1 \mathord{\left/{\vphantom{1 2}} \right. \kern-\nulldelimiterspace} 2}$&
20& 20& (0,10)& (0,10) \\
\hline
16 & $N(0,2)$ & $\chi_2^2 -2$& 20& 20& (0,10)& (0,10) \\
\hline
\end{tabular}
\caption{
\label{tab:list-sim-cases}
The simulated cases for the comparison of ANCOVA and methods on covariate-adjusted residuals.
$e_{ijk}$: error term;
$\stackrel{ind}{\sim}$: independently distributed as;
$n_i$: sample size for treatment level $i=1,2$.
$N\left({\mu,\sigma^2} \right)$ is the normal distribution with mean $\mu $ and variance $\sigma ^2$;
$\U\left({a,b} \right)$ is the uniform distribution with support $(a,b)$;
$DW\left({a,b,c} \right)$ is the double Weibull distribution
with location parameter $a,$ scale parameter $b,$ and shape parameter $c$;
$\be \left({a,b} \right)$ is the Beta distribution with shape parameters $a$ and $b$;
$\chi_2^2$ is the chi-square distribution with 2 {\it df};
$LN(a,b)$ is the log-normal distribution with location parameter $a$ and scale parameter $b$. }
\end{table}

\begin{table}[hbp]
\centering
\begin{tabular}{|c|c|c|c|c|c|c|c|c|c|c|}
\hline
&
\multicolumn{4}{|c|}{empirical sizes} &
\multicolumn{6}{|c|}{size comparison}  \\
\hline
Case & $\ah_1 $ & $\ah_2 $ & $\ah_3 $ & $\ah_4 $& (1,2)& (1,3)& (1,4)& (2,3)& (2,4)& (3,4) \\
\hline
1a&
.0531 & .0541$^{\ell}$ & .0540$^{\ell}$ & .0532 & $\approx $ & $\approx $ & $\approx $ & $\approx $ & $\approx $ & $\approx $ \\
\hline
1b & .0507 & .0493 & .0493 & .0510 & $\approx $ & $\approx $ & $\approx $ & $\approx $ & $\approx $ & $\approx $ \\
\hline
2a & .0581$^{\ell}$ & .0576$^{\ell}$ & .0546$^{\ell}$ & .0612$^{\ell}$ & $\approx $ & $\approx $ & $<$ & $\approx $ & $<$ & $<$ \\
\hline
2b & .0531 & .0515 & .0493 & .0630$^{\ell}$ & $\approx $ & $\approx $ & $<$ & $\approx $ & $<$ & $<$ \\
\hline
3a & .0606$^{\ell}$ & .0602$^{\ell}$ & .0567$^{\ell}$ & .0693$^{\ell}$ & $\approx $ & $\approx $ & $\approx $ & $\approx $ & $\approx $ & $<$ \\
\hline
4a & .0523 & .0525 & .0519 & .0511 & $\approx $ & $\approx $ & $\approx $ & $\approx $ & $\approx $ & $\approx $ \\
\hline
5a & .0490 & .0496 & .0499 & .0502 & $\approx $ & $\approx $ & $\approx $ & $\approx $ & $\approx $ & $\approx $ \\
\hline
6a & .0556$^{\ell}$ & .0024$^{c}$ & .0024$^{c}$ & .0033$^{c}$ & $>$ & $>$ & $>$ & $\approx $ & $\approx $ & $\approx $ \\
\hline
7a & .0465 & .0339$^{c}$ & .0337$^{c}$ & .0332$^{c}$ & $>$ & $>$ & $>$ & $\approx $ & $\approx $ & $\approx $ \\
\hline
8a & .0474 & .0437$^{c}$ & .0433$^{c}$ & .0440$^{c}$ & $\approx $ & $\approx $ & $\approx $ & $\approx $ & $\approx $ & $\approx $ \\
\hline
9a & .0485 & .0489 & .0484 & .0488 & $\approx $ & $\approx $ & $\approx $ & $\approx $ & $\approx $ & $\approx $ \\
\hline
10a & .0508 & .0505 & .0490 & .0595$^{\ell}$ & $\approx $ & $\approx $ & $\approx $ & $\approx $ & $\approx $ & $<$ \\
\hline
11a & .0522 & .0515 & .0511 & .0576$^{\ell}$ & $\approx $ & $\approx $ & $<$ & $\approx $ & $<$ & $<$ \\
\hline
12a & .0490 & .0494 & .0492 & .0491 & $\approx $ & $\approx $ & $\approx $ & $\approx $ & $\approx $ & $\approx $ \\
\hline
13a & .0486 & .0481 & .0480 & .0473 & $\approx $ & $\approx $ & $\approx $ & $\approx $ & $\approx $ & $\approx $ \\
\hline
14a & .0442$^{c}$ & .0435$^{c}$ & .0417$^{c}$ & .0451$^{c}$ & $\approx $ & $\approx $ & $\approx $ & $\approx $ & $\approx $ & $\approx $ \\
\hline
15a & .0383$^{c}$ & .0386$^{c}$ & .0357$^{c}$ & .0521 & $\approx $ & $\approx $ & $<$ & $\approx $ & $<$ & $<$ \\
\hline
16a & .0510 & .0514 & .0502 & .0701$^{\ell}$ & $\approx $ & $\approx $ & $<$ & $\approx $ & $<$ & $<$ \\
\hline
\end{tabular}
\caption{
\label{tab:emp-size}
The empirical sizes and size comparisons of ANCOVA and
methods on covariate-adjusted residuals for the 16 cases listed in Table \ref{tab:list-sim-cases}
based on 10000 Monte Carlo samples:
$\ah_i$: empirical size of method $i$;
$(i,j)$: empirical size comparison of method $i$ versus method $j$ for $i,j=1,2,3,4$ with $i \ne j$
where method $i=1$ is for ANCOVA, $i=2$ and $i=3$ are for ANOVA with and without HOV on covariate-adjusted residuals, respectively,
$i=4$ is for K-W test covariate-adjusted residuals.
$^{\ell}$( $^{c}$): Empirical size is significantly larger (smaller) than 0.05;
i.e., method is liberal (conservative).
$\approx$: Empirical sizes are not significantly different from each other;
i.e., methods do not differ in size.
$<$ ($>)$: Empirical size of the first method is significantly smaller
(larger) than the second.
}
\end{table}


\begin{table}[htbp]
\centering
\begin{tabular}{|c|c|c|c|c|c|c|}
\hline
&
\multicolumn{6}{|p{233pt}|}{Proportion of agreement}  \\
\hline
Case & $\ah_{1,2}$ & $\ah_{1,3}$ & $\ah_{1,4}$ & $\ah_{2,3}$ & $\ah_{2,4}$ & $\ah_{3,4}$ \\
\hline
1a & .0520$^n$ & .0519$^n$ & .0429$^s$ & .0540$^n$ & .0432$^s$ & .0431$^s$ \\
\hline
1b & .0490$^n$ & .0490$^n$ & .0415$^s$ & .0493$^n$ & .0413$^s$ & .0413$^s$ \\
\hline
2a & .0560$^n$ & .0545$^n$ & .0419$^s$ & .0546$^n$ & .0415$^s$ & .0405$^s$ \\
\hline
2b & .0513$^n$ & .0493$^n$ & .0383$^s$ & .0493$^n$ & .0377$^s$ & .0369$^s$ \\
\hline
3a & .0581$^n$ & .0565$^n$ & .0468$^s$ & .0567$^n$ & .0469$^s$ & .0453$^s$ \\
\hline
4a & .0507$^n$ & .0505$^n$ & .0382$^s$ & .0519$^n$ & .0380$^s$ & .0378$^s$ \\
\hline
5a & .0473$^n$ & .0389$^s$ & .0382$^s$ & .0396$^s$ & .0392$^s$ & .0388$^s$ \\
\hline
6a & .0024$^n$ & .0024$^n$ & .0033$^n$ & .0024$^n$ & .0015$^n$ & .0015$^n$ \\
\hline
7a & .0338$^n$ & .0336$^n$ & .0286$^s$ & .0337$^n$ & .0260$^s$ & .0260$^s$ \\
\hline
8a & .0426$^n$ & .0423$^n$ & .0346$^s$ & .0433$^n$ & .0340$^s$ & .0338$^s$ \\
\hline
9a & .0475$^n$ & .0473$^n$ & .0417$^s$ & .0484$^n$ & .0422$^s$ & .0420$^s$ \\
\hline
10a & .0498$^n$ & .0488$^n$ & .0420$^s$ & .0490$^n$ & .0421$^s$ & .0412$^s$ \\
\hline
11a & .0507$^n$ & .0504$^n$ & .0456$^s$ & .0511$^n$ & .0457$^s$ & .0454$^s$ \\
\hline
12a & .0477$^n$ & .0476$^n$ & .0378$^s$ & .0492$^n$ & .0383$^s$ & .0383$^s$ \\
\hline
13a & .0476$^n$ & .0476$^n$ & .0369$^s$ & .0480$^n$ & .0371$^s$ & .0371$^s$ \\
\hline
14a & .0425$^n$ & .0412$^n$ & .0274$^s$ & .0417$^n$ & .0275$^s$ & .0272$^s$ \\
\hline
15a & .0367$^n$ & .0355$^n$ & .0253$^s$ & .0357$^n$ & .0252$^s$ & .0246$^s$ \\
\hline
16a & .0497$^n$ & .0493$^n$ & .0394$^s$ & .0502$^n$ & .0392$^s$ & .0389$^s$ \\
\hline
\end{tabular}
\caption{
\label{tab:prop-agreement}
The proportion of agreement values for pairs of methods in
rejecting the null hypothesis for the 16 cases listed in Table \ref{tab:list-sim-cases}
based on 10000 Monte Carlo samples:
$\ah_{i,j}$: proportion of agreement between method $i$ and method $j$
in rejecting the null hypothesis for $i,j=1,2,3,4$ with $i \ne j$
where method labeling is as in Table \ref{tab:emp-size}.
$^n$: Proportion of agreement, $\ah_{i,j}$, is not significantly
different from the minimum of $\ah_i$ and $\ah_j$.
$^s$: Proportion of agreement, $\ah_{i,j}$, is significantly
smaller than the minimum of $\ah_i$ and $\ah_j$.
}
\end{table}


\begin{table}[htbp]
\centering
\begin{tabular}{|c|c|c|c|c|c|c|c|c|}
\hline
&
\multicolumn{8}{|c|}{cases}  \\
\hline
& 1a; 1b& 2a; 2b& 3a; 3b& 4a; 4b& 5a; 5b& 6a; 6b& 7a; 7b& 8a; 8b \\
\hline
$\kappa_1 $& 1.82; 1.30& 3.58; 2.38& 3.34; 2.38& 4.06; 3.06& 1.98; 1.42& 2.96; 2.20& 2.02; 1.40& 1.98; 1.32 \\
\hline
$\kappa_2 $& 1.82; 1.34& 3.58; 2.50& 3.34; 2.38& 4.40; 3.06& 2.06; 1.42& 5.36; 3.30& 2.28; 1.58& 2.02; 1.40 \\
\hline
$\kappa_3 $& 1.82; 1.34& 3.58; 2.50& 3.34; 2.38& 4.40; 3.06& 2.06; 1.42& 5.36; 3.30& 2.28; 1.58& 2.02; 1.40 \\
\hline
$\kappa_4 $& 1.90; 1.36& 3.80; 2.60& 3.36; 2.38& 4.38; 2.92& 2.06; 1.46& 7.04; 3.58& 2.70; 1.76& 2.04; 1.46 \\
\hline
& \multicolumn{8}{|c|}{cases}  \\
\hline
& 9a; 9b& 10a; 10b& 11a; 11b& 12a; 12b& 13a; 13b& 14a; 14b& 15a; 15b& 16a; 16b \\
\hline
$\kappa_1 $& 1.80; 1.30& 2.74; 1.98& 2.06; 1.44& 1.74; 1.18& 1.86; 1.22& 4.46; 2.78& 9.86; 5.58& 338; 2.34 \\
\hline
$\kappa_2 $& 1.80; 1.30& 2.74; 1.98& 2.06; 1.44& 1.74; 1.26& 1.86; 1.22& 4.46; 2.78& 9.86; 5.58& 3.42; 2.34 \\
\hline
$\kappa_3 $& 1.80; 1.30& 2.74; 1.98& 2.06; 1.44& 1.74; 1.26& 1.86; 1.22& 4.46; 2.78& 9.86; 5.58& 3.42; 2.34 \\
\hline
$\kappa_4 $& 2.02; 1.52& 3.20; 2.34& 2.34; 1.72& 2.02; 1.60& 1.98; 1.32& 4.10; 2.26& 3.66; 1.90& 3.62; 2.64 \\
\hline
\end{tabular}
\caption{
\label{tab:kappa-values}
The intercept difference values
at which the power estimates reach 1 for the 16 cases listed in Table \ref{tab:list-sim-cases}
based on 10000 Monte Carlo samples:
$\kappa_i$= intercept difference value at which power estimate of method $i$
reaches 1 for the first time for $i=1,2,3,4$
where method labeling is as in Table \ref{tab:emp-size}.
}
\end{table}

\newpage
\section{Figures}
\label{sec:figures}

\end{document}